\documentclass{article} 
\usepackage{emulateapj}
\usepackage{psfig}

\def\ledd{L_{\rm Edd}}

\newcommand\Teff{T_{\rm eff}}

\def\msun{{\,M_\odot}}

\def\refindent{\par\noindent\hangindent=3pc\hangafter=1 }
\def\aa#1#2#3{\refindent#1, A\&A, {\bf #2}, #3}

\def\apj#1#2#3{\refindent#1, {\it ApJ}, {\bf#2}, #3.}
\def\apjlett#1#2#3{\refindent#1, {\it ApJL}, {\bf #2}, #3.}

\def\mnras#1#2#3{\refindent#1, {\it MNRAS}, {\bf#2}, #3.}
\def\nature#1#2#3{\refindent#1, {\it Nature}, {\bf #2}, #3.}

\def\refpaper#1#2#3#4{\refindent#1, #2, #3, #4}
\def\refbook#1{\refindent#1}

\newcommand\fr{F_{\rm rad}}

\newcommand\pr{P_{\rm rad}}
\newcommand\pg{P_{\rm gas}}
\newcommand\ptot{P_{\rm tot}}

\newcommand\simlt{\lower.5ex\hbox{$\; \buildrel < \over \sim \;$}}
\newcommand\simgt{\lower.5ex\hbox{$\; \buildrel > \over \sim \;$}}

\newcommand\dm{\dot{m}}

\begin{document}

\title{Time-Dependent Disk Models for the Microquasar GRS~1915+105}

\author{Sergei Nayakshin\altaffilmark{1,2,3}, Saul
Rappaport\altaffilmark{4}, 
and Fulvio Melia\altaffilmark{5,6}}

% Notice that each of these authors has alternate affiliations, which
% are identified by the \altaffilmark after each name.  The actual
% alternate
% affiliation information is typeset in footnotes at the bottom of the
% first page, and the text itself is specified in \altaffiltext
% commands.
% There is a separate \altaffiltext for each alternate affiliation
% indicated above.

\altaffiltext{1}{Physics Department, The University of Arizona,
Tucson, AZ 85721.}
\altaffiltext{2}{LHEA, GSFC/NASA, Code 661, Greenbelt, MD,
20771.}
\altaffiltext{3}{National Research Council Associate.}
\altaffiltext{4}{Physics Department and 
Center for Space Research, MIT, Cambridge, MA 02139.}
\altaffiltext{5}{Physics Department and Steward Observatory, 
The University of Arizona, Tucson, AZ 85721.}
\altaffiltext{6}{Sir Thomas Lyle Fellow.}

% The abstract environment prints out the receipt and acceptance dates
% if they are relevant for the journal style.  For the aasms style, they
% will print out as horizontal rules for the editorial staff to type
% on, so long as the author does not include \received and \accepted
% commands.  This should not be done, since \received and \accepted dates
% are not known to the author.

\begin{abstract}
        During the past two years, the galactic black hole microquasar
GRS~1915+105 has exhibited a bewildering diversity of large amplitude,
chaotic variability in X-rays.  Although it is generally accepted that the
variability in this source results from an accretion disk instability, the
exact nature of the instability remains unknown. Here we investigate
different accretion disk models and viscosity prescriptions in order to
provide a basic explanation for the exotic temporal behavior in GRS~1915+105.

We discuss a range of possible accretion flow geometries.  Based on
the fact that the overall cycle times are very much longer than the
rise/fall time scales in GRS~1915, we rule out the geometry of advection
dominated accretion flow (ADAF) or a hot quasi-spherical region plus a cold
outer disk for this source.  A cold disk extending down to the last inner
stable orbit plus a hot corona above it, on the other hand, is allowed.

We thus concentrate on geometrically thin (though not necessarily
standard) Shakura-Sunyaev type disks (Shakura \& Sunyaev 1973;
hereafter SS73). We argue that X-ray observations clearly require a
quasi-stable accretion disk solution at high accretion rates where
radiation pressure begins to dominate, which excludes the standard
$\alpha$-viscosity prescription. To remedy this deficiency, we have
therefore devised a modified viscosity law that has a quasi-stable
upper branch, and we have developed a code to solve the time-dependent
equations to study such an accretion disk. Via numerical simulations,
we show that the model does account for several gross observational
features of GRS~1915+105, including its overall cyclic behavior on
time scales of $\sim$ 100 - 1000 s. On the other hand, the rise/fall
time scales are not short enough, no rapid oscillations on time scales
$\simlt$ 10 s emerge naturally from the model, and the computed
cycle-time dependence on the average luminosity is stronger than is
found in GRS~1915+105.

We then consider, and numerically test, several effects as a possible
explanation for the residual disagreement between the model and the
observations.  A hot corona with the energy input rate being a
function of the local cold disk state and a radius-dependent
$\alpha$-parameter do {\em not} appear to be promising in this
regard. However, a more elaborate model that includes the cold disk, a
corona, and plasma ejections from the inner disk region allows us to
reproduce several additional observed features of GRS~1915+105.  We
conclude that the most likely structure of the accretion flow in this
source is that of a cold disk with a modified viscosity prescription,
plus a corona that accounts for much of the X-ray emission, and
unsteady plasma ejections that occur when the luminosity of the source
is high. The disk is geometrically thin due to the fact that most of
the accretion power is drained by the corona and the jet.
\end{abstract}

\keywords{accretion disks, black holes, binary X-ray sources, 
microquasars}

\section{Introduction}

Compact X-ray sources exhibit a wide range of temporal variabilities
(from milliseconds to years). Perhaps none of these is as exotic and
diverse as the X-ray temporal variability observed from the black hole
microquasar GRS~1915+105 (Castro-Tirado, Brandt, \& Lund 1992; Greiner,
Morgan, \& Remillard 1996; Morgan, Remillard \& Greiner 1997; Muno,
Morgan \& Remillard 1999). This object is one of two known Galactic
X-ray sources that exhibit superluminal radio jets (Mirabel \&
Rodrigues 1994). The combination of relativistic constraints and radio
measurements at HI indicate that the source lies behind the
Sagittarius arm at a distance of 12.5 $\pm 1.5$ kpc (Mirabel \&
Rodrigues 1994).  Interstellar extinction limits optical/IR studies to
weak detections at wavelengths less than 1 micron (Mirabel et al.
1997).  The source is suspected to be a black hole binary because of
its spectral and temporal similarities with the other Galactic X-ray
source with superluminal radio jets, GRO~J1655-40 (Zhang et al 1994),
which has a binary mass function indicative of a black hole system
(Bailyn et al.  1995). Estimates for the mass of the compact object in
GRS~1915+105 range from 7 to 33 $\msun$. Even with the uncertainty in
distance, its peak X-ray luminosity is unusually high, i.e., $\simgt
10^{39}$ ergs/sec, which is around the Eddington luminosity for a
$\sim 7$ $\msun$ object.

In spite of several attempts it has proven especially illusive to
interpret the X-ray light curves of GRS1915. It is not yet clear that
even the basic time scales exhibited by the variability have been
successfully explained. Belloni et al. (1997a,b) accounted for the
observations with an empirical model in which the inner disk region
``disappears'' in the low count rate state, and is then replenished on
a viscous time scale. The parameters of their model are: the inner
disk radius, $R_{\rm in}$; the corresponding effective temperature of
the disk $T_{\rm in}$, and an ad-hoc non-thermal power law (which is
possibly produced in the disk corona). Although no detailed physical
model for the instability was given, very interesting patterns of
behavior for $R_{\rm in}$ and $T_{\rm in}$, as well as several other
observables, were deduced from the data, and the Shakura-Sunyaev
viscosity parameter was found to be unexpectedly low (which may mean
that the standard viscosity prescription is invalid for this
source). The rather small values of $R_{\rm in}$ found by these
authors can be used to discriminate between different models of the
accretion flow in GRS~1915 (see Appendix and \S \ref{sect:previous}).

Abramowicz, Chen \& Taam (1995) suggested a model for the low
frequency quasi-periodic oscillations (QPO) observed in selected X-ray
binaries, in which a corona above the {\em standard accretion} disk
leads to a mild oscillatory behavior. With some modifications, this
model could reasonably be expected to account for at least some of the
temporal variability in GRS~1915+105 as well (Taam, Chen \& Swank
1997).  However, it appears to us that the analysis of Abramowicz et
al.  (1995) and Taam et al. (1997) contains an error in the
heating/cooling equation for the disk which, when corrected,
constrains their model to have the same stability characteristics as a
standard Shakura-Sunyaev disk, and is therefore unlikely to explain
the GRS~1915+105 observations (see Appendix A).

We show more generally in Appendix \ref{sect:geometry} that neither a
hot central region, nor an advection-dominated flow, nor a ``slim''
accretion disk are compatible with the observations of
GRS~1915+105. (``Slim'' accretion disk theory was developed in the
most detail by Abramowicz et al. 1988; it is similar to a standard
thin Shakura-Sunyaev disk, except for the energy equation, which
incorporates the radial advection of energy into the black hole.) In
this paper, we attempt to undertake a more systematic study of the
variability patterns in GRS~1915+105 within the context of the ``cold
disk$+$hot corona'' picture. In \S \ref{sect:framework} we present a
general discussion that will guide us in our selection of a novel
(though somewhat ad-hoc) prescription for the viscosity in cases where
the radiation pressure is substantial (\S \ref{sect:prescription}).
The details of our numerical algorithm to solve the time-dependent
disk equations with the use of this new viscosity prescription are
given in \S \ref{sect:code}.

\begin{figure*}
\centerline{\psfig{file=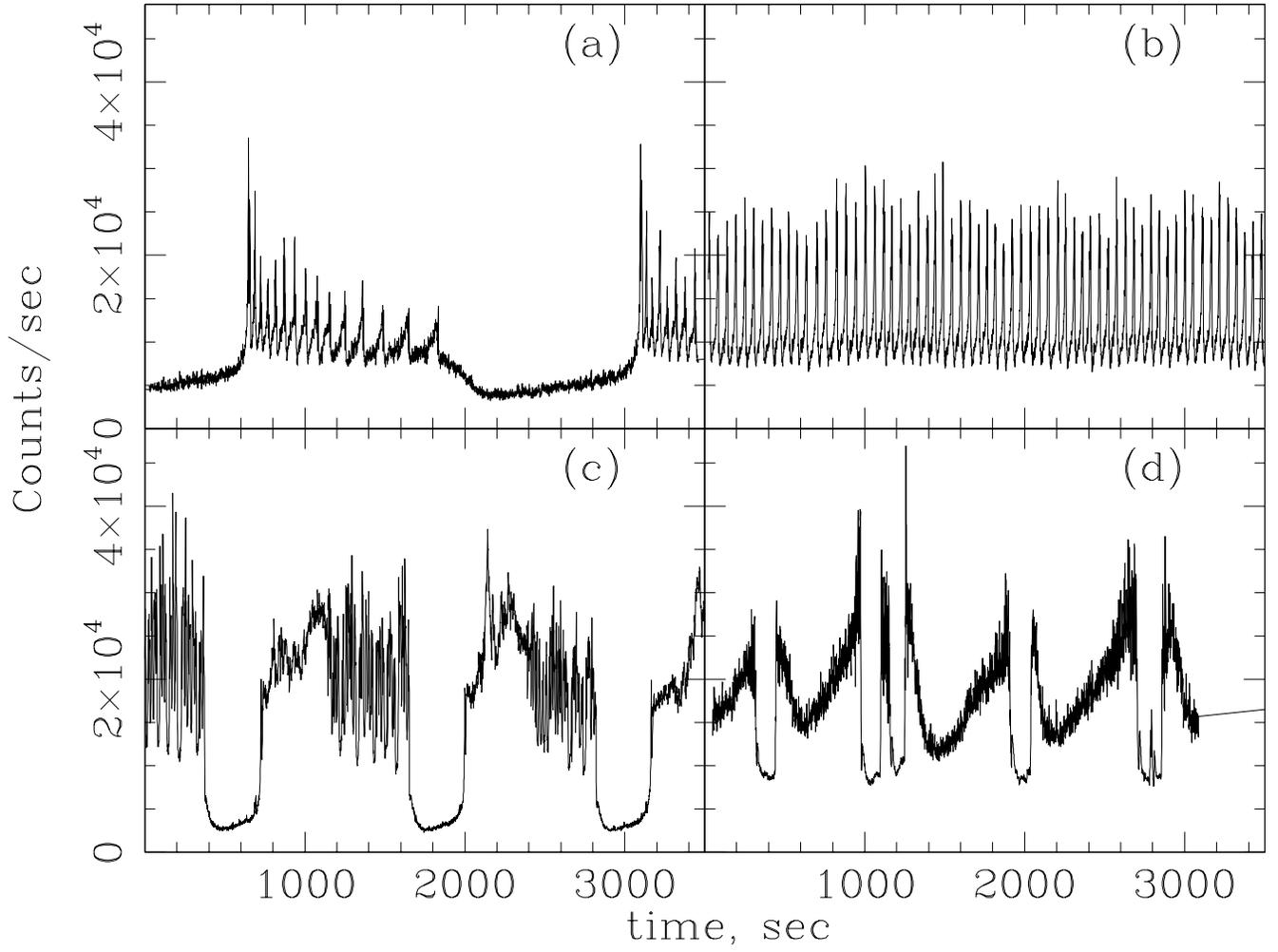,width=.95\textwidth,angle=-90}}
\caption{Selected segments of the X-ray light curve of GRS~1915+105,
demonstrating the broad range of variability patterns in this unusual
source. The data were obtained from the public archive of the Rossi
X-ray Timing Explorer (RXTE).}
\label{fig:data}
\end{figure*}

In \S \ref{sect:results}, we present the results of our time-dependent
disk calculations. The calculated light curves are found to agree
qualitatively with many observational features of GRS~1915+105. In
particular, the characteristic cycle times and duty cycles are in
reasonable agreement with the observations. Moreover, the trend in the
cycle time with the average accretion rate, $\dm$ has the correct
sense. However, there are important disagreements as well. We
therefore introduce a more elaborate model in \S \ref{sect:jet},
where, in accordance with the observations (e.g., Mirabel \& Rodrigues
1994), the inner disk is allowed to expel some of its energy in the
form of a non-steady jet.  We assume that the ejected energy is not
observed in X-rays, but rather that it ultimately produces radio
emission. We show that this more elaborate model agrees with the
GRS~1915+105 observations much better, perhaps indicating that we are
finally developing a zeroth order understanding of the geometry and
the most important processes in this enigmatic source.  In \S
\ref{sect:previous} we discuss our results in the light of the earlier
work on GRS~1915, and in \S \ref{sect:discussion} we summarize our
conclusions.

\section{Limit Cycle Behavior in GRS~1915+105}
\label{sect:framework}

Figure 1 shows four examples of typical X-ray light curves from GRS
1915 obtained with the RXTE satellite (see, e.g., Morgan, Remillard,
\& Greiner 1997). It appears that the source undergoes a
limit-cycle type of instability; the cycle times in panels (a) -- (d)
are $\sim$ 2400, 60, 1200, and 800 sec, respectively.  Within the
cycles shown in panels (c) and (d) there appear yet other
quasi-regular oscillations of still shorter time scale.  The shortest
time scale over which a substantial change in the X-ray flux occurs is
$\sim$ 5 sec.  We will associate the longer cycles with the viscous
time scales of the inner accretion disk, whereas the more rapid
behavior will be related to the thermal time scale.  The X-ray
spectrum of GRS~1915+105 varies systematically with X-ray intensity;
usually the spectral hardness is strongly correlated (or
anti-correlated) with intensity (e.g., Belloni et al. 1997a,b; Taam et
al. 1997; Muno et al. 1999). More specifically, the spectrum is
typically found to be composed of a multi-temperature disk component,
hereafter the ``thermal component", and a power-law component with a
varying spectral index.  Finally, the source exhibits a wide range of
QPOs with central frequencies in the range of 0.01 - 10 Hz. The
amplitude and frequency of the QPOs appear to be strongly correlated
with the spectral state of GRS~1915+105 (Chen et al. 1997; Muno et
al. 1999).

\begin{figure*}[t]
\centerline{\psfig{file=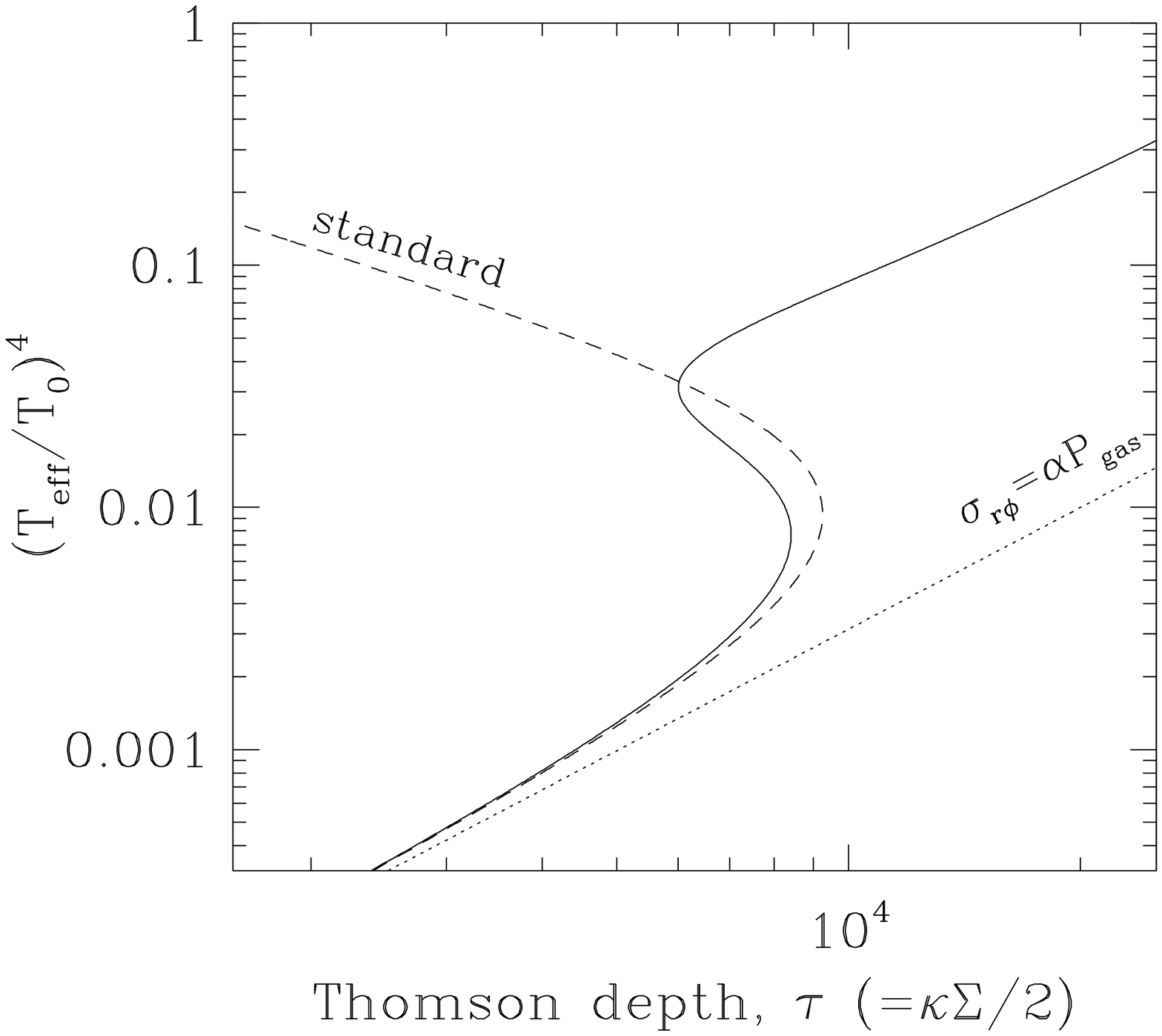,width=.6\textwidth}}
\caption{Effective temperature versus the Thomson optical depth at a
fixed radial distance $R_0$ for accretion disks with the standard (dashed
curve), Lightman-Eardley (dotted), and our modified viscosity
prescription (solid curve). $T_0$ is defined as the effective
temperature of the disk at $R_0$ for the case where the disk radiates
at the Eddington luminosity (e.g., $\dm =1$). The lower bend in the
S-curve occurs due to the radiation pressure starting to dominate
within the disk, while the higher bend is introduced by requiring the
viscosity law to follow the Lightman-Eardley form above some value for
$\pr/\pg$.}
\label{fig:scurve}
\end{figure*}

The idea of using a modified viscosity law to test time-dependent
phenomena in accretion disks around black holes is not new (e.g., Taam
\& Lin 1984; Chen \& Taam 1994; and further references below in this
section). But let us examine in global terms how the various disk
configurations relate to the behavior of GRS~1915+105 in
particular. In geometrically thin disks the advection of energy is not
important. The viscous time scale is much longer than both the thermal
and hydrostatic time scales (see Frank et al. 1992), unless
$\alpha\sim 1$ which is unlikely for this source (Belloni et
al. 1997b). In this context, then, the equations for vertical
hydrostatic equilibrium and energy balance between viscous heating and
radiative losses should be valid.

It is customary and instructive to plot the disk effective temperature
$\Teff$ versus the column density $\Sigma$ to investigate the
stability properties of disks. In the one zone limit, the equation for
hydrostatic equilibrium in the vertical direction is
\begin{equation} 
\ptot \equiv \pg + \pr = {G M\Sigma H\over 2 R^3 }\,,
\label{eq1} 
\end{equation}
where $M$ is the mass of the black hole, $H$ is the scale height and
$R$ is the radial distance from the central object. The gas pressure
is given by $\pg = 2 \rho\, m_p^{-1} kT$ (for hydrogen rich material), 
where $T$ is the mid-plane
temperature, $m_p$ is the proton mass, and the radiation pressure is
$\pr = a T^4/3$.  The energy balance equation is given by
\begin{equation} 
\fr \simeq {c\pr\over \tau} = {3\over 2} \sigma_{r\phi}\, \Omega H\,,
\label{eq2} 
\end{equation}
where $\fr$ is the vertical radiation flux, $\tau$ is the optical
depth of the disk, $\tau\equiv \kappa \Sigma/2$, and $\kappa$ is the
radiative opacity (assumed to be dominated by electron scattering
opacity), $\sigma_{r\phi}$ is the stress tensor (see, e.g., Shapiro 
\& Teukolsky, \S 14.5), and $\Omega$ is the Keplerian angular 
frequency. As we said earlier, in the standard Shakura-Sunyaev 
formulation, the stress tensor is proportional to the total pressure,
\begin{equation}
\sigma_{r\phi}=
\alpha \ptot\,{\rm .}
\label{ss}
\end{equation}

We can now combine Equations (\ref{eq1}) -- (\ref{ss}) to eliminate
the variable H and thereby derive an expression to generate the
$\Teff$--$\Sigma$ curve for any given viscosity. In Figure
(\ref{fig:scurve}) we show several illustrative $\Teff$--$\Sigma$
curves to help motivate our selection of $\alpha$. The dashed curve
represents the track that the standard Shakura-Sunyaev theory with
$\alpha =$ const produces. Such a model is thermally and viscously
unstable when the slope of the $\Teff$--$\Sigma$ curve becomes
negative (e.g., Frank et al. 1992), which might be a desirable result
in the view of the instabilities observed in GRS~1915+105. However,
the observed light curves require a quasi-stable accretion mode for
the high-luminosity state as well, which is {\it not predicted} by the
standard theory (see also the end of App. B). This suggestion (already
discussed by Belloni et al. 1997a,b and Taam et al. 1997) is motivated
by the fact that in some cases the upper luminosity state of the
source lasts even longer than the low state, which is presumably the
stable gas-pressure-dominated branch.  A $\Teff$-$\Sigma$ relation
that has a stable high-luminosity state is shown in Figure
(\ref{fig:scurve}, solid curve). It has the characteristic
``S-shape'', familiar from the studies of the thermal ionization
instability in accretion disks (e.g., Meyer \& Meyer-Hofmeister 1981,
Bath \& Pringle 1982, Cannizzo, Chen \& Livio 1995, and Frank et
al. 1992, \S 5.8) or from studies of radiation-pressure driven
instabilities by, e.g., Taam \& Lin (1984) and Chen \& Taam (1994). It
is well known that parts of an S-curve that have a negative slope are
unstable. In short, when the $\Teff$-$\Sigma$ relation produces an
S-curve, the unstable region of the disk oscillates between the two
stable branches of the solution, which gives rise to the high and low
luminosity states.

So far, we have not considered flows where advection is important.  As
ADAFs (Advection Dominated Accretion Flows) will be discussed in \S
\ref{sect:adaf}, here we concentrate on the slim accretion disks
(Abramowicz et al. 1988). These flows have a stable high $\dm$
solution even with the standard viscosity prescription due to the
additional cooling (energy transport) via advection. However, the
S-curves produced by these models have the upper stable branch for
$\dm\gtrsim 1$ only (see, e.g., see Fig. 2 in Chen \& Taam 1993). In
other words, if one were to position the upper stable branch of the
slim accretion disks on our Figure (\ref{fig:scurve}), then it would
look similar to the corresponding stable branch of the solid curve in
the figure, except that it would be positioned higher by a factor of
about $100$ (at a given $\Sigma$). Thus, if one assumes that an
unstable part of the disk makes a transition from the lower bend in
the S-curve upwards to the stable branch, and that $\Sigma$ remains
approximately constant during this process, then the accretion rate in
the inner disk must increase by as much a factor of a few hundred to a
thousand to reach the stable advective regime.

While this may not be a problem in terms of the luminosity output,
since most of the energy is advected inwards and the luminosity of the
disk saturates at $L\simeq L_{\rm Edd}$ (see Fig. 1 in Szuszkiewicz,
Malkan \& Abramowicz 1996), the light curves produced will always be
very spiky and the duty cycle of the high state is always much smaller
than the values of $\sim 0.5$ that are often seen in GRS~1915+105. The
physical cause of the spiky behavior is the vast difference between
the accretion rates in the high and low states. As we will discuss
later in \S \ref{sect:results}, an outburst starts because ``too
much'' mass (column density $\Sigma$) has been accumulated above the
maximum possible stable value of $\Sigma_{\rm max}$ (i.e., $\Sigma$ at
the lower bend of the S-curve in Figure \ref{fig:scurve}). If the high
stable state accretion rate is very much larger than the low state
accretion rate, it then takes only a very short time in the high state
to remove this excess mass from the inner disk.  In other words, it
would not be possible to maintain mass conservation if the accretion
rate persisted at the high state for very long.  This is indeed seen
from the temperature curves computed for slim disks by Szuszkiewicz \&
Miller (1998, see their Figures 7-9).  Therefore, we believe that slim
disks are not likely to adequately account for the behavior of
GRS~1915+105, at least not if one seeks a theory capable of explaining
all of the variability in this source.

\section{The Basic Model}\label{sect:basic}

\subsection{Time-Dependent Equations For Disk-Corona Models}
\label{sect:code}

The geometry in our unstable disk model is that of a standard
Shakura-Sunyaev configuration (except for the viscosity law) overlayed
with a hot corona. Here we describe the set of equations that we use
to follow the temporal evolution of this disk; these equations are
applicable to any local viscosity law. A discussion of the viscosity
itself is deferred until \S \ref{sect:prescription}.  The standard
Euler equations for conservation of mass and angular momentum in the
disk (see, e.g., Frank et al. 1992) can be combined in the usual way
to yield the equation describing the evolution of the surface density:
\begin{equation}
{\partial \Sigma \over \partial t} = {3\over R}\,
{\partial \over \partial R} \left\{R^{1/2}{\partial \over \partial R}
\left[\nu\Sigma R^{1/2}\right] \right\}\,,
\label{eq4}
\end{equation}
where all of the variables, except for $\nu$, were defined in the
previous section.  Here $\nu$ is the viscosity that is related to
$\alpha$ by $\alpha \ptot$ = ${3\over 2} \nu \Omega \rho$.

The time-dependent energy equation includes the heating and cooling
terms of Equation (\ref{eq2}), but must also be able to take into
account large radial gradients of temperature, and therefore includes
a number of additional terms.  The form of the energy equation that we
use follows the formalism of Abramowicz et al. (1995) with some
modifications, and is given by

\begin{eqnarray}
\ptot H \;{4-3\beta\over \Gamma_3 -1}\;\Bigg[ \left( {\partial \ln T
\over \partial t} \, + \,v_R {\partial \ln T
\over \partial R}\right) \nonumber \\
- \left(\Gamma_3-1\right) \,
\left({\partial \ln \Sigma\over \partial t} \, +\, v_R
{\partial \ln \Sigma\over \partial R} - {\partial H\over
\partial t}\right) \Bigg] \nonumber \\
= F^{+} - F^{-} - {2\over R}\, {\partial(R F_R H)\over
\partial R}  + J  ,
\label{eq5}
\end{eqnarray}
where $\beta =
1/(1+\xi)$, $\xi = \pr/\pg$ is the ratio of the radiation to the
gas pressure, $\gamma$ is the ratio of specific heats ($\gamma = 5/3$)
and $\Gamma_3$ is given in Abramowicz et al. (1995). The radial
velocity $v_R$ is given by the standard expression (Eq. 5.7 of
Frank et al. 1992):
\begin{equation}
v_R = -{3\over \Sigma R^{1/2}}	\,{\partial\over \partial R}
\left[\nu\Sigma R^{1/2}\right].
\label{vr}
\end{equation}

The terms on the left hand side of Equation (\ref{eq5}) represent the
full time derivative (e.g., $\partial/\partial t + v_R
\partial/\partial R$) of the gas entropy, while the terms on the right
are the viscous heating, the energy flux in the vertical direction,
the diffusion of radiation in the radial direction, and the viscous
diffusion of thermal energy.  Following Cannizzo (1993; and references
cited therein), we take $J = 2 c_p
\nu (\Sigma/R) [\partial(R \partial T/\partial R)/ \partial R]$ to be
the radial energy flux carried by viscous thermal diffusion, where
$c_p$ is the specific heat at constant pressure.  $F^{+}$ is the
accretion disk heating rate per unit area, and is given by 
\begin{equation}
F^{+} = {9\over 4} \nu \Sigma R\,\Omega_K^2   .
\label{heatr}
\end{equation}
The radiation flux in the radial direction is
\begin{equation}
F_R = - 2 {c\pr\over \tau_T}\, H {\partial \ln T\over \partial R}  .
\label{raddif}
\end{equation}
The overall cooling rate (larger than simply that from radiative
diffusion) in the vertical direction is given by
\begin{equation}
F^{-} = {c\pr\over \tau_T} (1-f)^{-1}   ,
\label{fvert}
\end{equation}
where $0\leq f < 1$ is the fraction of the power that is transferred
to the disk surface by mechanisms other than the usual radiation
diffusion (cf. Abramowicz et al. 1995, Svensson \& Zdziarski 1994).
Equations (\ref{heatr}) and (\ref{fvert}) differ from those used by
Abramowicz et al. (1995) for reasons that are detailed in Appendix B.
We note that in the most general sense, the quantity $(1-f)^{-1}$
should be thought of as a factor correcting the vertical energy
transport in the standard accretion disk theory. In our formulation of
the problem, it is computationally irrelevant whether this fraction of
energy is deposited in the corona and radiated as a non-thermal
power-law component, or it contributes to the blackbody disk flux,
since we are concerned only with the bolometric luminosity of the disk
for now. The latter case could be realized if this additional energy
transport were to deposit its energy just below the photosphere of the
disk. Of course, if our model produces light curves that are in
reasonable agreement with those of GRS~1915+105, then a future study,
one that would carefully delineate the different spectral components,
will be warranted.

Equations (\ref{eq4}) and (\ref{eq5}), together with the equation of
hydrostatic equilibrium (\ref{eq1}), yield a closed set of coupled,
time-dependent equations for evolving the variables T and $\Sigma$.
For solving these equations we employ a simple explicit scheme with a
variable time step. The time step is chosen to be always a fraction of
the smallest thermal time scale in the disk (i.e., in its inner part).
We also use a fixed grid in $R$. The spacing is uniform in $R^{1/2}$,
and we take 100 to 300 radial bins. We will often refer to the radial
coordinate in a dimensionless form, i.e., $r\equiv R/R_g$, where $R_g=
2 GM/c^2$ is the Schwarzschild radius.  The outer boundary of the disk
is chosen at a large radius, $r_{\rm max}\sim$ few hundred to a
thousand, such that the disk at $r\simgt r_{\rm max}/3$ is always
gas-pressure dominated. Thus, the outer boundary conditions are given
by $T=T_{\rm ss}$ and $\Sigma =
\Sigma_{\rm ss}$ for $r=r_{\rm max}$, where $T_{\rm ss}$ and
$\Sigma_{\rm ss}$ are the temperature and the column density in the
Shakura-Sunyaev formulation at that radius for the given $f$, $\dm$
and $\alpha = \alpha_0$, where $\alpha_0$ is the viscosity parameter
for a gas-dominated disk (see below). The inner boundary of the disk
is fixed at $r=3$, and the boundary conditions there are $T=0$ and
$\Sigma=0$.

\subsection{Modified Viscosity Law}\label{sect:prescription}

In this section we present our modified viscosity law that we anticipate 
will account for the unstable behavior of GRS~1915+105. 
The Shakura-Sunyaev viscosity prescription postulates that
the viscous stress tensor $\sigma_{r\phi}$ is given by Equation
(\ref{ss}), where the parameter $\alpha$ is defined as
\begin{equation}
\alpha \simeq (v_t l_t/c_s H) + (B^2/4\pi
P_{\rm tot}) {\rm ,}  
\label{ssvis}
\end{equation}
and where $v_t$ and $l_t$ are the turbulent eddy velocity and
scale-length, respectively, $c_s$ is the sound speed, and $B^2/4\pi$
is the volume averaged magnetic field energy density. Obviously, this
approach is useful only when $\alpha$ remains approximately constant,
i.e., independent of the local thermodynamic variables in the disk,
which is probably the case in a number of situations. However,
whenever the physical state of the accreting gas changes considerably,
it is hard to see why $\alpha$ would not change. One confirmation of
this is the well-known thermal (ionization) instability of accretion
disks in dwarf nova systems (e.g., Meyer \& Meyer-Hofmeister 1981;
Bath \& Pringle 1982; Cannizzo, Chen \& Livio 1995; and Frank et
al. 1992; \S 5.8). It has been shown in this case that the
$\alpha$-parameter needs to be larger on the hot stable branch than it
is on the low branch of the solution. Some more recent work suggests
that, in addition, $\alpha$ should have a power-law dependence on
$(H/R)$, e.g., $\alpha \propto (H/R)^{3/2}$, in order to explain the
observed exponential luminosity decline of the dwarf nova outbursts
(Cannizzo et al. 1995; Vishniac \& Wheeler 1996).

Further, based on theoretical considerations, different authors have
adopted several variants on the prescription for $\sigma_{r\phi}$ when
$\pr\geq \pg$. For example, Lightman \& Eardley (1974; hereafter
LE74), and Stella \& Rosner (1984; hereafter SR84) suggested that
\begin{equation}
\sigma_{r\phi} = \alpha_0\pg\,,
\label{le}
\end{equation}
where $\alpha_0$ is a constant. Their physical reasoning was based on
the expectation that chaotic magnetic fields should be limited by the
gas pressure only.  In addition, Sakimoto \& Coronitti (1989;
hereafter SC89) showed that even if one assumes that magnetic fields
are effectively generated by radiation-dominated disks, magnetic
buoyancy quickly expels these fields from the disks. Since the disk
viscosity may be produced to a large extent by magnetic fields, this
result implies that the $\alpha$-parameter must decrease when the
radiation pressure becomes dominant. An attractive feature of this
viscosity prescription (Eq. \ref{le}) is the fact that it is stable
even when $\pr\geq \pg$, as indicated by the positive slope generated
by this model for the track in $\Teff$--$\Sigma$ parameter space (see
the dotted curve in Fig. \ref{fig:scurve}). At the same time, however,
this viscosity prescription (to which we will refer as the
Lightman-Eardley prescription) has no unstable region at all, and thus
fails to reproduce the unstable behavior observed in GRS~1915+105.

Of course, the conclusions of LE74, SR84 and SC89 should be regarded
as qualitative, at best, since it is not yet feasible to describe the
turbulence and magnetic fields immersed in a radiation-dominated fluid
in a quantitative and model-independent way. In this paper, we adopt
the view that it may still be possible to couple the radiation to the
particles through collisions and thereby allow the radiation pressure
to contribute to the viscosity when $\pr/\pg$ is not too great. For
example, SR84 argued that the energy density in the chaotic magnetic
fields cannot exceed the particle energy density, since it is the
latter that generates the magnetic fields in the first place. However,
as long as $\pr/\pg < \alpha^{-1}$, the magnetic field pressure does
not exceed the gas pressure (because then $B^2/8\pi \sim
\alpha \pr < \pg$), so that the SR84 arguments are superfluous.

\begin{figure*}[t]
\centerline{\psfig{file=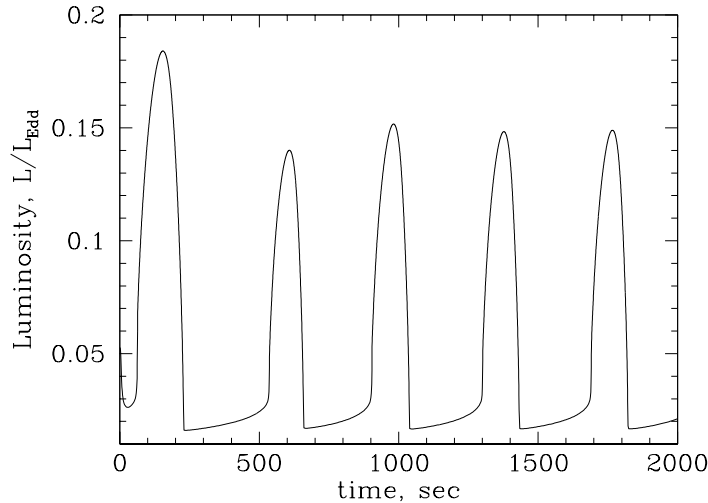,width=.5
\textwidth,angle=-90}}
\caption{The light curve for an unstable disk with parameters $f=0$
and $\dm = 0.05$.}
\label{fig:feqo}
\end{figure*}

To cast our suggestion into a mathematical form, we have adopted a
viscosity prescription intermediate between those of Shakura-Sunyaev
and Lightman-Eardley. Our viscosity law is proportional to the total
pressure for small to moderate values of $\xi = \pr/\pg$, while for
large values of $\xi$, it becomes proportional to the gas pressure. We
have devised the following simple formula for the viscosity
prescription that smoothly joins these two limits, such that $\alpha$
is given by
\begin{equation} 
\alpha = \alpha_0 {1 + \xi/\xi_0\over 1 +	
(\xi/\xi_0)^2} \;,
\label{eq3} 
\end{equation} 
where $\xi_0$ is an adjustable parameter. The S-curve plotted in
Figure (\ref{fig:scurve}) was generated using this viscosity
prescription with $\xi_0 =8$. 

\section{Results}\label{sect:results}

\subsection{Accretion Disk Evolution Through One Cycle}
\label{sect:evol}

There are essentially 5 free parameters of the model that need to be
specified and/or systematically varied to ascertain the sensitivity of
the results to their value: ({\it i}) the black hole mass, $M$, ({\it
ii}) the average accretion rate, $\dm$, ({\it iii}) the viscosity
parameter, $\alpha_0$, for the gas-dominated region, ({\it iv})
$\xi_0$, the critical ratio $\pr/\pg$, and ({\it v}) $f$, the fraction
of the luminosity transported vertically by mechanisms other than
radiative diffusion. We adopt a mass of $10\msun$, which is typical of
the measured masses of galactic black hole candidates in binary
systems (see, e.g., Barret et al. 1996). For purposes of illustration,
we chose a value of $\alpha_0$ equal to $0.01$, since it leads to
viscous time scales comparable to those observed. Similarly, we
selected a value of $\xi_0 = 8$, in part because it produces a
reasonable S-curve. However, we checked a range of other values as
well (from $\xi_0=3$ to $\xi_0=20$), and found that $\xi = 8$
reproduces the GRS~1915+105 observations most closely.

\begin{figure*}[t]
\centerline{\psfig{file=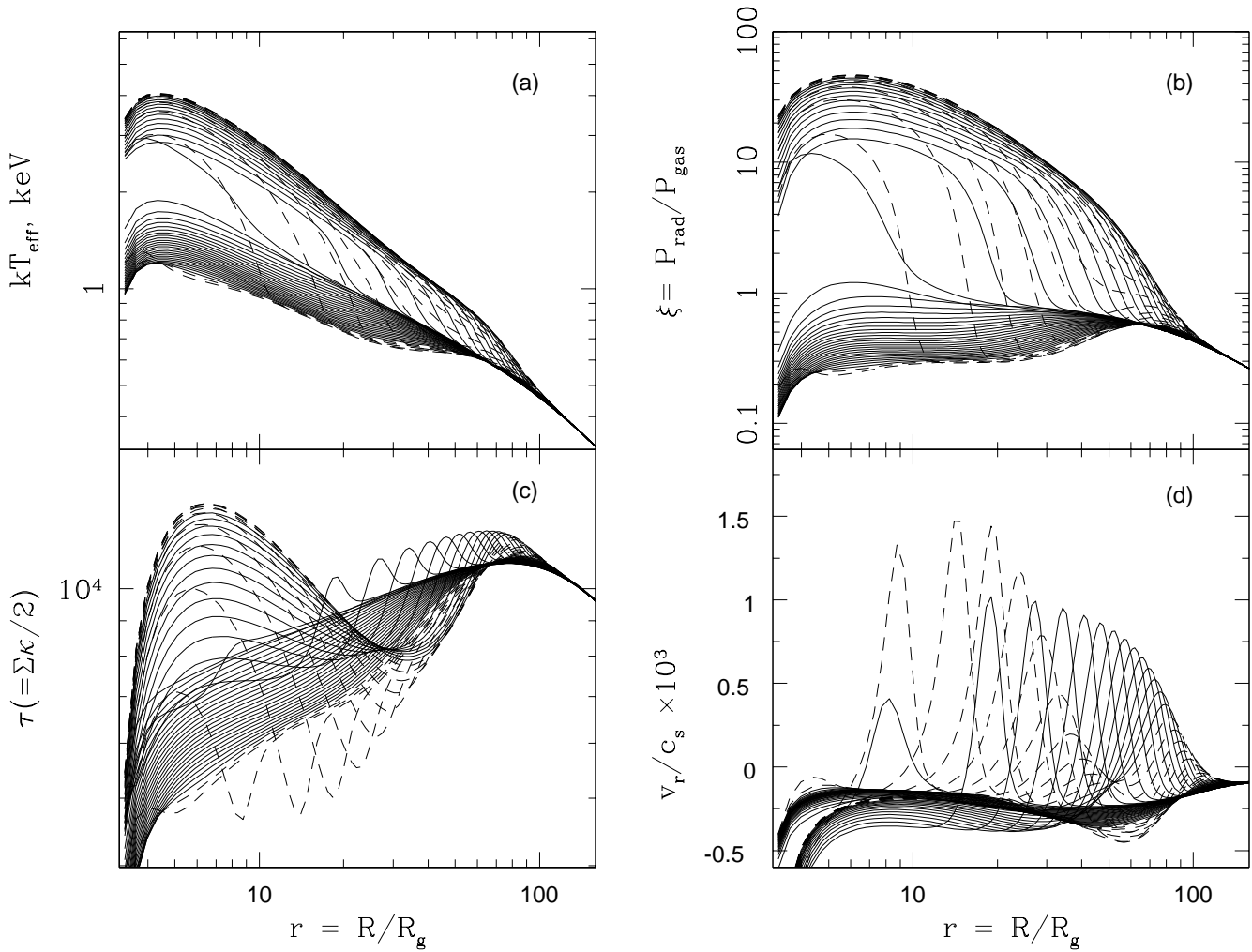,width=.95
\textwidth,angle=-90}}
\caption{The evolution of (a) the disk temperature, (b) the 
ratio of radiation to gas pressure, $\xi$, (c) the Thomson optical
depth, and (d) the ratio of radial flow velocity to the sound speed
throughout one cycle for $f=0$, $\dm = 0.05$. Each curve is a snapshot
of the parameter versus radius $R$, taken after
each $\sim 7.5$ sec from about 670 to 1030 seconds in
Fig. (\ref{fig:feqo}). The solid curves correspond to times when the
luminosity increases with time (e.g, during the low state and during
the rise to the maximum), whereas the dashed curves correspond to
times when the luminosity decreases.}
\label{fig:temp_tau_distr}
\end{figure*}

We present a computed light curve and variations of the disk
parameters through one complete cycle in Figures (\ref{fig:feqo}) and
(\ref{fig:temp_tau_distr}), respectively, for one set of illustrative
parameters: $\dm = 0.05$ and $f=0$. In Figure
(\ref{fig:temp_tau_distr}), panels (a), (b), (c) and (d) show the
evolution of the disk effective temperature $T_{\rm eff}$, the ratio
of radiation to gas pressure ($\xi$), the disk Thomson optical depth
$\tau$ and the radial flow velocity $v_r$. Note that since we are
currently concerned only with the luminosity integrated over all
photon energies, we define $T_{\rm eff}$ to include the emissivity
from {\em both} the corona and the disk, i.e., $\sigma_B T_{\rm
eff}^4\equiv F^{-}$, where $F^{-}$ is given by Equation
(\ref{fvert}). In each panel, the series of curves are for temporal
increments of $\sim 7.5$ sec.

The curve with the smallest luminosity $L$ (where $L = \int 2\pi\, R\,
\sigma_B T_{\rm eff}^4\, dR\,$) in the upper left panel of Figure
(\ref{fig:temp_tau_distr}) corresponds to the state of the disk right
after the end of an outburst.  An outburst starts in the innermost
region of the disk, when a few zones at the smallest radii quickly
make a transition to the upper stable branch of the solution (high
state). Although this is not obvious from
Fig. (\ref{fig:temp_tau_distr}), the viscosity of the high state is
larger than it is in the low state, because $\nu = \alpha c_s H$, and
so even though $\alpha$ decreases, the increase in $c_s$ and $H$ leads
to a larger $\nu$. The larger viscosity allows the gas to dispose of
its angular momentum faster, thus allowing a faster inflow of the gas
into the black hole. Because of angular momentum conservation, the
angular momentum of the gas that plummets into the black hole is
transferred to larger $R$, where it produces an excess of angular
momentum, and thus some matter actually flows to larger $R$, which is
seen in panel (d) of Figure (\ref{fig:temp_tau_distr}) as positive
spikes in the radial velocity distribution. A heating wave is
initiated and propagates from the inner disk to larger $R$
(Fig. \ref{fig:temp_tau_distr}a). This wave is often referred to as
either a `density wave' or a `transition wave'. As the wave propagates
outward, the material on the inside of the wave becomes hot and shifts
into the high viscosity state. It then rapidly loses its angular
momentum and is transferred into the innermost disk, where the
material continuously plummets into the hole. Since the innermost
stable region is a ``bottle-neck'' for the accreting gas, some
material builds up there, which explains the bump in $\Sigma$ in that
region as seen in Figure (\ref{fig:temp_tau_distr}c) during the
outburst.

When the density wave reaches $r\sim 100$, the instability saturates,
since the disk in that region is always gas-pressure dominated (for
the chosen $\dm$ and $f$).  As the density wave dies away, the
outermost unstable regions are cleared of the excess mass and cool
down. They are now in the low viscosity state, and thus the rate of
mass inflow in the innermost region of the disk drops. An inward
propagating cooling wave is initiated and the outburst dies out as the
last excess mass in the innermost disk sinks into the hole. Finally, a
new cycle begins with the accumulation of new excess mass (since the
local accretion rate is smaller than the average rate).

\subsection{Light curves}\label{sect:lc}

We now systematically explore the disk behavior with the new viscosity
prescription, in order to eventually compare the results with the
gross temporal behavior of GRS~1915+105.  We adopted an energy
transport fraction $f$ = 0.9.
As with the other parameters, we also tested a wide range of values of
$f$, and found that $f= 0.9$ is the most appropriate due to the following
considerations. The threshold for the onset of the instability is
about $\dm_0 =$ 0.26 for $f =$ 0.9, and scales as:
\begin{equation}
\dm_0 = 0.02\, \left[1-f\right]^{-9/8}
\label{dmo}
\end{equation}
(see Svensson \& Zdziarski 1994 and note that their definition of
$\dm$ differs from ours by a factor that accounts for the standard
disk radiative efficiency, i.e., $0.06$). It seems clear
observationally that $\dm_0$ should be larger by a factor of at least
$5$ than the value of $\dm = 0.02$ corresponding to the transition
from the gas- to radiation pressure-dominated regime in the standard
theory. Indeed, if we assume that the maximum luminosity of $\sim
2\times 10^{39}$ erg sec$^{-1}$ observed in GRS~1915+105 corresponds
to the Eddington luminosity for this source, then the instability
seems to exist only for $\dm \gtrsim 0.1-0.2$. In addition, Galactic
Black Hole Candidates (GBHCs) with a lower persistent luminosity, or
weaker transient behavior, have not shown such violent instabilities
as GRS~1915+105 does, and yet many of them are as bright as $0.1\ledd$
(see, e.g., Barret, McClintok \& Grindlay 1996). We will discuss the
parameter $f$ further in \S \ref{sect:gross}.

\begin{figure*}[t]
\centerline{\psfig{file=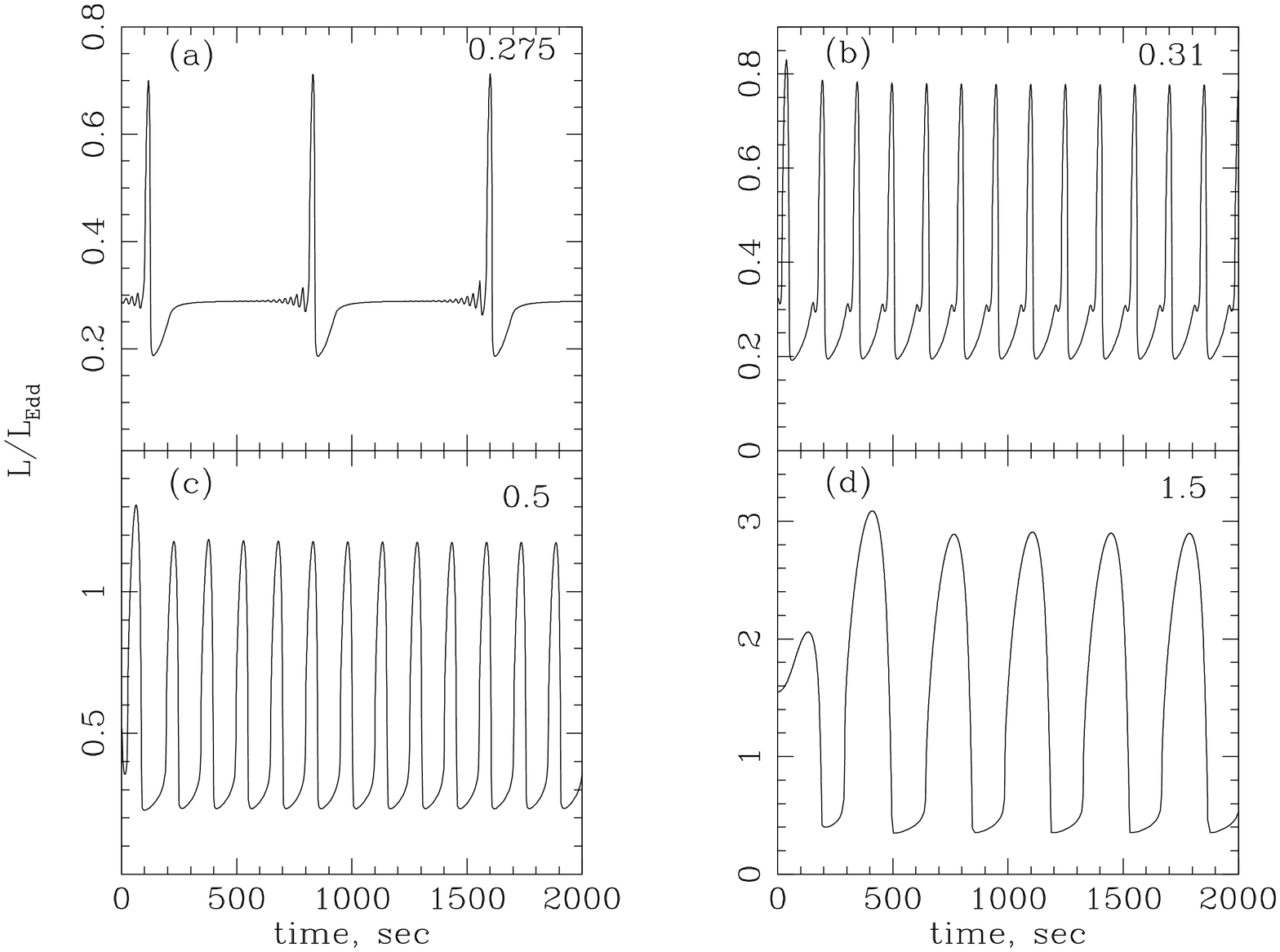,width=.97\textwidth,angle=-90}}
\caption{The light curves for an unstable accretion disk with a
viscosity prescription given by Equation (\ref{eq3}). The coronal
dissipation fraction $f$ is set to $0.9$, $\alpha_0 = 0.01$, and
$\xi_0 = 8$ for all tests. The dimensionless accretion rate $\dm$ is
shown in the right upper corner of each figure.}
\label{fig:lcurves}
\end{figure*}

In Figure (\ref{fig:lcurves}) we show the evolution of the integrated
disk luminosity as a function of time for the nominal values of the
parameters listed above and for $\dm=$ $0.275$, 0.31, 0.5, and 1.5.
In agreement with Equation (\ref{dmo}), we find that for values of
$\dm$ below 0.26 the disk is stable, because there is no
radiation-pressure dominated zone in the disk. We also find, in
general, that as the value of $\dm$ increases, the duty cycle (i.e.,
the fraction of time that the source spends in the high state) increases,
from about 5\% at $\dm=$ 0.26, to approximately 50\% when $\dm=$ 1.5.
Similarly, the ratio of maximum to minimum luminosity through the
cycle grows with increasing $\dm$.

The time for the disk to complete one of its cycles (defined as the
cycle time) also increases with $\dm$, except for very near the
minimum value of $\dm=$ 0.26 where the instability first sets in; here
the cycle time actually decreases with increasing $\dm$. The reason is
that there seems to exist a ``critical excess mass'' in the inner disk
region, such that the instability will appear only when the mass in
the disk exceeds this value. The rate for building up the excess mass
is $\dm-\dm_0$, where $\dm_0$ is the maximum stable accretion rate
given by Equation (\ref{dmo}), and therefore a small increase in $\dm$
above $\dm_0$ can bring about a substantial decrease in the cycle
time. For larger accretion rates, the cycle time starts to increase
with increasing accretion rate, because a larger region of the disk is
unstable, and it takes longer to clear the excess mass during the
outburst.

\subsection{Comparison to Observations of GRS~1915+105}\label{sect:th_and_obs}

We will now compare our results shown in Figure (\ref{fig:lcurves})
with observations of GRS~1915+105. We should acknowledge at the start
that it seems a daunting task to try to explain with a single model
the diverse, unstable behavior exhibited by this source, but we expect
to benefit from a comparison of even the gross properties of our model
with the observations.

\subsubsection{Gross Features}\label{sect:gross}

Panel (a) of Figure (\ref{fig:lcurves}) is similar to Figure
(\ref{fig:data}a) in that in both figures there exists a long phase of
mass accumulation (the low state). Furthermore, an examination of the
GRS~1915+105 data reveals that the cyclic limiting behavior disappears
when the average count rate is lower than $\sim 5 \times 10^3$
counts/sec, so that there indeed exists a threshold accretion rate
below which the instability does not operate. In simulations, the
spiky nature of the outburst is explained by the fact that only a very
narrow region in $R$ within the disk is unstable, and it takes very
little time to get rid of the excess mass in that region. The
differences with Fig. (1a) are mainly due to the fact that an outburst
in the data does not seem to completely clear the excess mass, and
that more oscillations follow with a gradually declining amplitude,
whereas in the simulations the hot state persists until all the excess
mass is swallowed by the black hole.

Panel (b) of Fig. (\ref{fig:data}) is also somewhat similar to panels
(b) and (c) of the simulations (Fig. \ref{fig:lcurves}). The
progression of time scales and the average luminosity (i.e., roughly
speaking, the count rate) from pattern (a) to (b) in
Fig. (\ref{fig:data}) is similar to that from panel (a) to (b) in Fig.
(\ref{fig:lcurves}) as well. The physical reason for the cycle time
getting shorter with increasing $\dm$ is that the excess mass can
build up faster for larger accretion rates.

Next, panel (c) in Figure (\ref{fig:data}) is most interesting from
the point of view that the high state is long lasting and is clearly
stable at least for the first half. Panel (d) of the simulations can
account for some of the properties of this variability pattern. In
particular, the duty cycle increased in both the simulations and the
data as the accretion rate increased. The cycle time did not increase
nearly enough in the simulations, however. It could be made longer by
choosing a smaller $\alpha_0$, but then the time scales in all the
other panels would increase as well, which does not appear acceptable.

Also, a general trend seen in the first three panels (a-c) in both the
data and the simulations is that the minimum of each light curve shows
a slow gradual rise in luminosity before the instability sets in.
This is a clear indication of the disk accumulating mass until it
reaches a global instability.

We note that panel (d) of Figure (\ref{fig:data}) seems to be rather
different from the other three variability examples. If the high state
is again defined as the one with larger count rates, then a
peculiarity of panel (d) is that the count rate first decreases and
then increases by the end of a high state episode. This is the
opposite of what is seen for the other states, such as that shown in
panel (c) of Fig. (1). From a theoretical point of view, this is a
highly significant observation.  In all of our simulations, we
observed that as the outburst progresses, the amount of mass in the
inner disk region builds up because of the greater and greater inflow
of mass from larger radii. Once in a given (high or low) stable state,
the local disk luminosity is proportional to the local column density
$\Sigma$, which can be seen from Figure (\ref{fig:scurve}). Therefore,
in an outburst, the luminosity decreases only after the outer disk
cools down and the influx of mass stops, leading to a decrease in
$\Sigma$ in the inner disk.  Thus, the outburst profile in the
simulations is such that the second time derivative $d^2 L/d t^2$ is
always negative, not positive as seen in panel (d) of Figure 1. We do
not see a clear explanation for this disagreement in the context of
the current model, but we will show that there might be a natural
cause for this phenomenon if a jet is allowed (\S
\ref{sect:jet}).

Finally, we note that in a number of the panels in Figure
(\ref{fig:lcurves}), the first (and sometimes also the second) peak in
the light curve is of slightly different intensity or duration than
the peaks that come later in the sequence.  This is simply a
consequence of the disk adjusting to its quasi-steady cyclic pattern
after starting from the imposed initial conditions that are given by
the Shakura-Sunyaev solution corresponding to the mean value of $\dm$.

Let us now discus the rather extreme value for $f$ invoked here, which
requires that as much as $90\%$ of the accretion power is carried out
of the disk by processes other than the usual radiative diffusion. If
this process is magnetic buoyancy or some other MHD process heating
the corona, then there appears to be a contradiction with the
observations of GRS~1915+105.  The blackbody component in the spectrum
of this source is generally small in the low state, but it can be
dominant in the outburst state (Muno, Morgan \& Remillard 1999). One
could argue here that as much as $80-90\%$ of the X-rays emitted by
the corona towards the disk may be reprocessed into the soft disk
radiation, (e.g., Magdziarz \& Zdziarski 1995) if the
reflection/reprocessing takes place in a {\em neutral medium} (neutral
in the sense of high-Z elements like oxygen and iron, that are
important for the reflection spectrum). Thus, in principle, the
blackbody flux could exceed that in the non-thermal component by the
ratio $\sim (2-f)/f$ (see Eqs. 1--5 in Haardt \& Maraschi 1991 with
their parameters $a\simeq 0$ and $\eta\simeq 1/2$).  The rather high
inferred disk temperature in GRS~1915+105, e.g., $kT\sim 2$ keV
(Belloni et al. 1997a), rules out the possibility of neutral
reflection, however. As found by Nayakshin \& Dove (1998) and
Nayakshin (1998), the integrated albedo of the reflected spectrum in
the case of a strongly ionized disk can be much higher than the
standard 0.1-0.2 of the neutral reflector (i.e., $a\simgt 0.5$ in
Eqs. 1--5 of Haardt \& Maraschi 1991), and thus it seems difficult to
have $f$ as large as 0.9 given the large amount of soft power observed
from the disk.

A possible way out of this dilemma is that the high value of $f$ does
not necessarily represent the energy flux from the disk into the
corona. Indeed, Equation (\ref{fvert}) states that the energy
transport out of the disk may be $(1-f)^{-1}$ times faster than that
given by the standard diffusion of radiation.  Convection of energy in
the vertical direction (e.g., Bisnovatiy-Kogan \& Blinnikov 1977;
Goldman \& Wandel 1995; and references therein) is one physical
mechanism that can speed up the transfer of energy out of the disk.
MHD waves dissipating their energy before they reach the corona could
be another. In addition, there is no proof that the vertical averaging
procedure (i.e., a one zone approximation) used in the standard
accretion disk theory does not lead to a substantial underestimation
of the vertical radiation flux out of the disk. For example, if
$F_{\rm rad} = (c\pr/\tau_d)\,
\zeta$, where $\zeta\sim$ few, then $\dm_0$ becomes $\dm_0 = 0.02
\zeta^{9/8}$ (see SZ94), and thus one may have $\dm_0$ as large
as that observed in GRS~1915+105 due to a faster or an additional disk
cooling mechanism rather than due to a transfer of most of the disk
power into the corona.

\subsubsection{Rapid Flickering}\label{sect:flickerring}

It is notable that panel (c) of Fig. (\ref{fig:data}) shows rapid
chaotic oscillations as fast as $\sim$ tens of seconds at the end of
the high state, whereas our simulations do not show a similar
behavior.  Furthermore, note that the rise/fall time scales are
shorter in the data than they are in our model. It appears that the
heating of the unstable region in GRS1915 happens on a time scale that
is very much shorter than the cycle time. By contrast, the outermost
part of the unstable region of the simulated disk becomes unstable
only after $\sim 1/2$ the duration of the hot phase, that is of the
order of the disk viscous time for the high luminosity cases.

One explanation here could be that these fast oscillations are failed
attempts by the disk to make a state transition from the high to the
low state. Further, during these fast oscillations the disk is always
brighter than it is in the low state, which could be interpreted as an
indication that only a part of the disk (most likely the innermost
region) in GRS~1915+105 takes part in these rapid oscillations, whereas
the rest of the unstable region is still on the upper stable branch of
the S-curve. For this to be true, the inner disk must be able to
decouple to some extent from the rest of the unstable disk, and be
variable on a much shorter time scale than the outer disk.

\begin{figure*}[t]
\centerline{\psfig{file=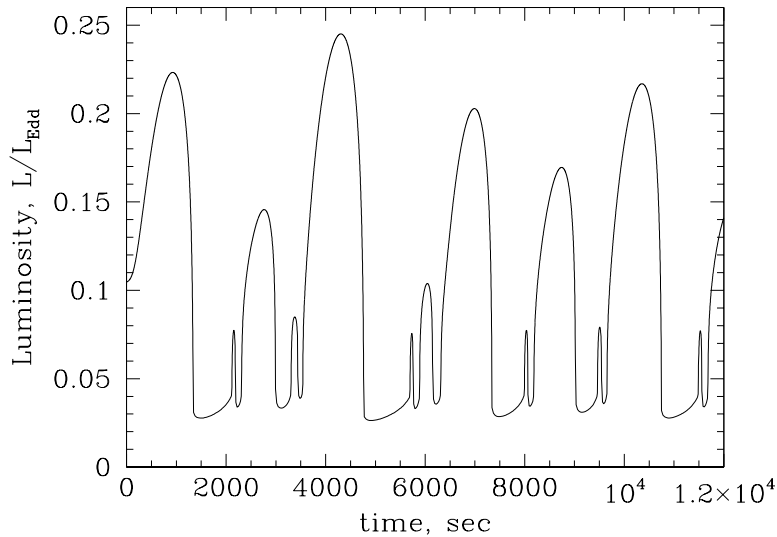,width=.55
\textwidth,angle=-90}}
\caption{The effects of varying the $\alpha$-parameter with radius,
according to $\alpha(\xi, r) = \alpha(\xi) \{0.1 + [1 +
(r/8)^3]^{-1}\}$. This effectively gives two different viscous time
scales, which is clearly seen in the resulting light curves. See text
for discussion.}
\label{fig:variable_alpha}
\end{figure*}

To test this idea, we have carried out several simulations in which
the $\alpha$-parameter is a decreasing function of radius. Our hope is
that in this case, since $\alpha$ in the inner disk is larger than
that in the outer disk, the time scale for oscillations in the inner
disk may be much shorter than the overall viscous time, which has
the dependence $t_v\propto \alpha^{-1}$ (e.g., Frank et al. 1992). In Figure
(\ref{fig:variable_alpha}) we show one such simulation, in which the
viscosity parameter was chosen to be $\alpha(\xi, r) = \alpha(\xi)
[0.1 + \{1 + (r/8)^3\}^{-1}]$. This functional form allows
the $\alpha$-parameter to be roughly $\alpha(\xi)$ in the inner disk
region ($R \leq 8 R_g$), and to be $\sim 0.1
\alpha(\xi)$ for $R > 8 R_g$. The presence of these two time scales is
obvious just from a perusal of the resulting light curves.
Oscillations of the inner disk produce the precursor seen before each
major outburst in Fig. (\ref{fig:variable_alpha}). Its relative
magnitude is small because the amount of excess mass stored in the
inner disk is small compared to that in the outer disk. Furthermore,
in the simulations the inner disk decouples from the outer one only at
the beginning of the outburst, not at the end as is seen in the
GRS~1915+105 data. This is due to the fact that once the outer disk
makes a transition to the high state, the large mass supply to the
inner disk forces the latter to go into the high state as well and
remain there. In summary , varying the $\alpha$-parameter with radial
distance does not appear to be a viable explanation for the rapid
chaotic oscillations seen in the data which are superposed on the more
regular long time scale disk evolution.

An alternative way of explaining the fast oscillations could have been
provided if the heating/cooling fronts stall and are reflected back as
cooling/heating fronts. This behavior was observed in simulations of
the classical thermal ionization disk instability by Cannizzo (1993,
see text above his Equation 5), where it happened to be an unwanted
result. However, we have not been able to see such stalled transition
fronts (except for the case presented in
Fig. \ref{fig:variable_alpha}), which should not be too surprising,
due to the fact that the underlying physics of the ionization
instability and the one explored in this paper are vastly different,
and one does not expect a direct correspondence between these two
instabilities.

Although we will not present any light curves, we should mention that
we have also attempted to allow the fraction $f$ to be a function of
$\xi$, since the transition from $\xi<1$ to $\xi > 1$ means a
substantial change in the physical conditions in the disk. Our hope
was that the freedom in choosing $f(\xi)$ might help to reproduce the
disk flickering. However, all our attempts in this regard (with $f$
decreasing or increasing across the transition in $\xi$) have been
unsuccessful.

\section{Plasma Ejection From the Inner Disk}\label{sect:jet}

Although our model is able to reproduce a number of the general trends
seen in the light curve of GRS~1915+105, the rise/fall time scales are
always shorter in the data than they are in the simulations. The fast
rises and falls are perhaps the reason why the profile of an outburst
in some of the actual light curves of GRS~1915+105 (e.g., panel (c) of
Fig. \ref{fig:data}) is reminiscent of a square-wave like shape rather
than the rounded shapes that an outburst has in the simulations (see
Fig. \ref{fig:lcurves}d). The difference between our simulations and
the data appears important enough to us to require the addition of a
new feature to our model and test whether it can bring the theory
closer to the observations.

The following considerations have guided us in choosing the additional
ingredient in our model. The speed with which the hot state propagates
outward in the accretion disk is equal to the speed of the transition
wave, which was studied analytically by, e.g., Meyer (1984) and
numerically by Menou, Hameury \& Stehle (1998). Both of these studies
find that the propagation speed of the transition wave is $v_t\simeq
\alpha c_s$.  The time scale for the transition front to traverse an
unstable region of size R is $t_t \sim R/\alpha c_s = (R/H) t_{\rm
th}$, i.e., the transitions in the disk actually take $R/H$ times
longer than the thermal time scale of the outermost part of the
unstable disk region. This is the reason why the rises and falls take
too long in the simulations compared to the data. 

However, notice that during the initial stage of the transition from
the low to the high state, the luminosity rises quite rapidly in the
simulations (see Fig
\ref{fig:lcurves}d), at a rate probably as fast as that seen in the data.
Now, if some physical process were to ``cut'' the simulated light
curves at, say, $L = L_{\rm Edd}$, so that anything emitted by the
disk above this value {\em is not seen} by the observer, then the
outburst profile would be much more consistent with the data.

\begin{figure*}[t]
\centerline{\psfig{file=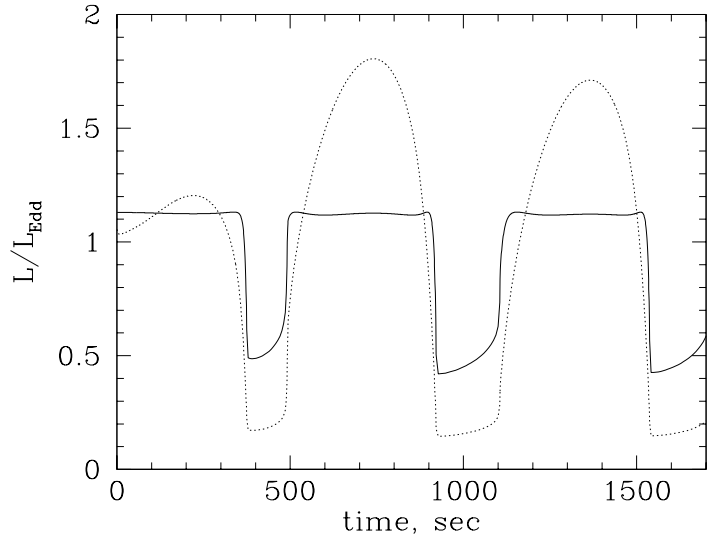,width=.5\textwidth,angle=-90}}
\caption{The light curve for
the disk, corona and jet model. The solid curve shows the X-ray
luminosity, whereas the dashed curve shows the overall disk power
(i.e., X-ray luminosity plus the energy ejected in the jet) scaled
down by a factor of 3 for clarity of presentation. The accretion rate
used for this test is $\dm = 3$, and the fraction of the power
emerging as X-rays depends on the overall disk power $\Lambda
\equiv L_{\rm t}/L_{\rm Edd}$ as $\eta_x = [1  + 
(\Lambda/2)^2]^{-1}$. The viscosity prescription and other parameters
of the model are the same as those used in Fig. (\ref{fig:lcurves}).}
\label{fig:steady1}
\end{figure*}

To find a process able to ``cut'' the light curve, we should recall
that powerful plasma ejections are known to occur in GRS~1915+105
(e.g., Mirabel and Rodriguez 1994). Recently, Eikenberry et
al. (1998), Mirabel et al. (1997) and Fender \& Pooley (1998) have
shown that there exists a strong link between the X-ray emission and
emission in the infrared and radio frequencies.  The radio and
infrared emission is attributed to plasma ejection events taking place
in the innermost part of the disk. The minimum power deposited at the
base of the jet (that may consist of individual ejection events) was
found (e.g., Gliozzi, Bodo \& Ghisellini 1999) to be as high as
$L_j\sim $ few $\times 10^{40}$ erg sec$^{-1}$ for the event first
discovered by Mirabel and Rodriguez (1994). During other observations
(e.g., Eikenberry et al. 1998), the radio fluxes were a factor of
$\sim 10$ lower than the ones observed by Mirabel \& Rodriguez (1994),
which still may require $\sim $ few $\times 10^{39}$ erg sec$^{-1}$
ejected into the jet, which is as large as the largest X-ray
luminosities produced by the source. Thus, if one were to model these
plasma ejections in the framework of our model, one would have to
allow for a significant portion of the accretion {\em energy} to be
diverted into the jet.

From here on, we will therefore assume that the total accretion disk
power $L_t$ consists of two parts: the first is the X-ray luminosity
$L_x$, that is equal to the sum of the thermal disk and non-thermal
coronal luminosities; and the second part $L_j$ is the ``jet power'',
i.e., the energy ejected into the jet. For simplicity we further
assume that the jet power is not seen in X-rays, and appears only
in the radio or other non-X-ray wavebands. As before, the locally
produced total disk power is given by Equation (\ref{fvert}), where we
will again suppose that $f=0.9$.  The jet luminosity is then
\begin{equation}
L_j = 2\times2\pi R_g^2 \int_{3}^{r_{max}} r dr \, f\,\eta_j F^{-}
\label{lj}
\end{equation}
where $0\leq \eta_j\leq 1$ is the fraction of power that escapes from
the disk into the jet. This parameter ($\eta_j$) may be a constant, or
it may be a function of local disk conditions or a function of some
global variable, e.g., the total disk power $L_t$. Since the
importance of advection relative to the radiative cooling is roughly
given by the ratio $(H/R)^2$, and since according to our discussion in
Appendix \ref{sect:geometry} $H/R\ll 1$ in this source, the advection
of energy is not important in our model. Thus, the observed X-ray
power $L_x$ will consist from the disk power (fraction $1-f$ of the
total disk power) and the coronal luminosity which is given by an
equation analogous to Equation (\ref{lj}) except with a parameter
$\eta_x = 1 -\eta_j$ instead of $\eta_j$ inside the integral. To
eliminate any ambiguities of our approach, we note that the accretion
power is divided among the disk, the corona, and the jet as $1-f$, $f
\eta_x$ and $f \eta_j $, respectively.

We should also mention that the amount of mass carried away by the jet
is small compared to that accreted into the black hole, so we neglect
the former in the mass conservation (eq. \ref{eq4}). The point here is
rather simple: if the ultimate source of the jet power is the
underlying accretion disk, then to produce one relativistic proton in
the jet, many protons in the disk must transfer their gravitational
energy (which is small compared with their rest mass) to that one jet
proton.  These protons will therefore have to sink into the black hole
in order to send one proton into the jet. More precisely, the mass
outflow rate is $\dot{M}_j\simlt L_j/\gamma_j c^2$, where
$\gamma_j\sim$ few is the terminal bulk Lorenz factor of the jet. This
is to be compared with the accretion rate through the disk, that is
equal to $\dot{M}_d = L_t/\epsilon c^2 $, where $\epsilon$ is the
``radiative'' efficiency of the disk and $L_t$ is the total disk
power. Accordingly,
\begin{equation}
{\dot{M}_j\over \dot{M}_d} \simlt \epsilon \gamma_j^{-1}\; {L_j\over
L_t} \simlt 0.1\;,
\label{mratio}
\end{equation}
since $\epsilon \simlt 1/6$. This estimate is quite conservative. For
example, if the jet power is dominated by magnetic fields or by
electron-positron pairs, then the mass carried away in the jet is even
smaller. A similar consideration allows us to neglect the angular
momentum outflow from the disk into the jet. We point out however that
the latter approximation depends on the model of the jet; if the jet
particles actually gain a considerable amount of the angular momentum
as they are streaming away from the disk, then the torque exerted by
the jet on the disk may become non-negligible.  We plan to explore
these effects in future work.

\begin{figure*}[t]
\centerline{\psfig{file=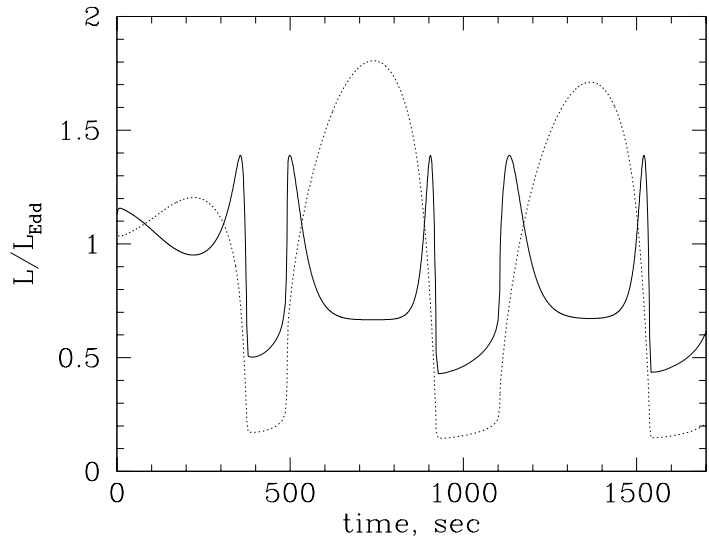,width=.5\textwidth,angle=-90}}
\caption{Same as Fig. (\ref{fig:steady1}), except the fraction of power
going into X-rays is now given by $\eta_x = [1 + (\Lambda/3)^2 +
(\Lambda/3)^6]^{-1}$. Notice that during the high state, the X-ray
light curves exhibit local minima rather than maxima.}
\label{fig:steady2}
\end{figure*}

Eikenberry et al. (1998) found that the ejection events were absent in
the low state and started at the onset and during the high state.  We
thus will accept a parameterization of $\eta_j$ such that it is close
to zero when the disk is in the low state, and it is relatively large
(but still smaller than unity, of course) when the disk is in the high
state. We now explore three different illustrative prescriptions for
the function $\eta_j$ to show the possible effects of the disk power
being diverted into a jet.

\subsection{Test 1}\label{sect:test1}

Let us assume that the fraction $\eta_j$ is described by the following
(somewhat arbitrarily chosen) function:
\begin{equation}
\eta_j = 1 - {1\over 1 + (\Lambda/2)^2 }\;,
\label{fj1}
\end{equation}
where $ \Lambda\equiv L_t/L_{\rm Edd}$ is the total power of the
source in terms of its Eddington luminosity. This equation is
qualitatively reasonable since one expects particles to be ejected
from the disk when the luminosity $L$ approaches (and then exceeds)
the Eddington limit. While the radiation may not be the ultimate
driving power of the jet, the radiation pressure can cause more matter
to be ejected and possibly produce stronger jets.  We show in Figure
(\ref{fig:steady1}) the resulting X-ray light curve (solid line) and
the total disk luminosity $L_t$ (represented by a dashed curve; scaled
down by the factor of 3 to fit the figure) for an accretion rate $\dm
= 3$. Note that the overall luminosity curve is the same as we would
have obtained with our basic model for the same choice of $f$ and
$\dm$, but with no jet ejections (see \S\S \ref{sect:basic} \&
\ref{sect:results}).

The shape of the outburst indeed becomes more like a square-wave, and
thus the rises and falls appear to be sharper than they were for the
model with $\eta_j=0$, thus making it possible to reproduce this
feature of the observed light-curves (in particular, panel (c) of Fig.
\ref{fig:data}).

\subsection{Test 2}\label{sect:test2}

As a second example, we test the following prescription for the jet
power:
\begin{equation}
\eta_j = 1 - {1\over 1 + (\Lambda/3)^2 +
(\Lambda/3)^6 }\;.
\label{fj}
\end{equation}
(This prescription, in contrast to the one given by eq. [\ref{fj1}],
was chosen among several that we tested to reproduce the panels (c)
and (d) of Fig. [\ref{fig:data}] simultaneously as described in \S
\ref{sect:test3}). The corresponding X-ray light curve and the total 
luminosity of the system are shown in Figure (\ref{fig:steady2}). One
notices that the profile of the outburst in X-rays is now inverted
with respect to the actual disk power. When the disk produces most of
the energy output, the X-ray light curve actually has a local minimum,
since most of the energy is ejected into the jet during that
time. This effect may be responsible for the ``strange'' shape of the
outburst seen in panel (d) of Fig (\ref{fig:data}).

\subsection{Test 3}\label{sect:test3}

As a final example we present a sequence of light curves produced by
our model with a fixed prescription for the jet energy fraction
$\eta_j$ and with the accretion rate being the only parameter that is
varied. The functional dependence of $\eta_j$ is again given by eq.
(\ref{fj}).  Further, motivated by the fact that the X-ray emission
can fluctuate wildly when it is in the high state (see Fig. 1),
whereas it does not fluctuate as much in the low luminosity state, we
allow the fraction $f$ carried away from the disk other than by the
usual radiation diffusion to fluctuate around some mean. Numerically,
the $f$-factor will now contain a {\em random} variable part $f_1 =
0.03$, such that $ f_1\ll f_0$:
\begin{equation}
f = f_0 + f_1 \,\mu\, g(r)\,
{\rm \Lambda^2\over 4 + \Lambda^2}\;,
\label{ft}
\end{equation}
where $f_0$ is the constant part fixed at the previously selected
value of $0.9$, $\mu$ is a random number distributed uniformly between
-1 and +1 (a new value of $\mu$ is randomly chosen every 3 seconds),
and $g(r)\equiv r_1^4/(r^4+r_1^4)$ is a function of radius $r$ such
that it is unity in the inner disk and it approaches zero for $r> r_1
= 30$. The latter radial dependence is introduced simply to avoid
complications with the outer boundary condition at $r = 10^3$, where
we set the disk structure to be that given by the gas-dominated
Shakura-Sunyaev configuration with a constant $f=f_0$ (see \S
\ref{sect:code}). The shape of the function $g(r)$ is almost
completely irrelevant for the resulting light curves as long as $r_1$
is large, since most of the disk power is liberated in the inner disk
region. Finally, although the second term on the right hand side of
the Equation (\ref{ft}) appears to be complicated and model-dependent,
its presence is not required and serves only to make the light curves
look somewhat more random. None of the final conclusions of this paper
depend on the particular choice for the random part in $f$.

\begin{figure*}[t]
\centerline{\psfig{file=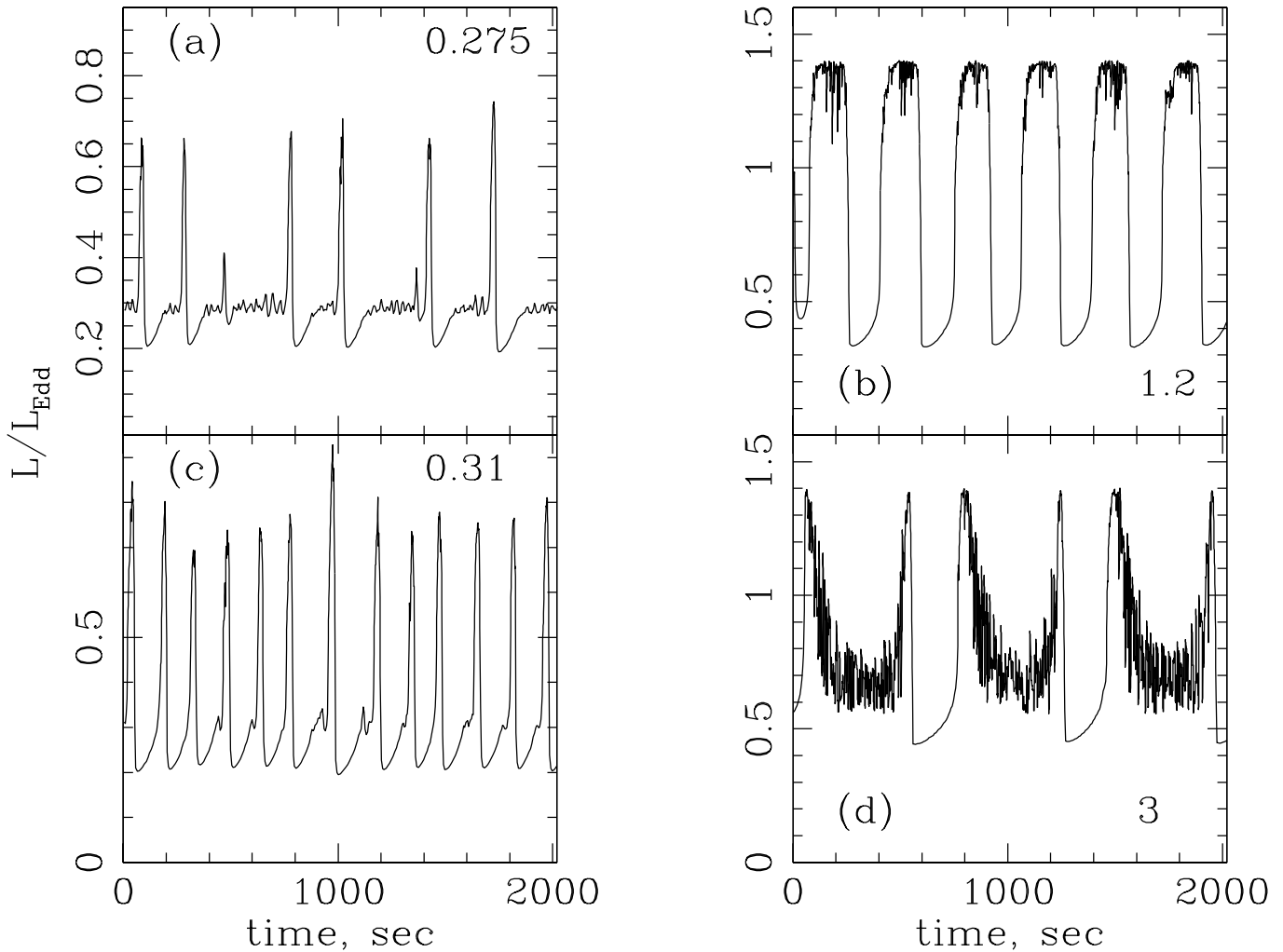,width=.97
\textwidth,angle=-90}}
\caption{The light curves for the unstable accretion disk with a
viscosity prescription given by Equation (\ref{eq3}), and with the
fluctuating corona and plasma ejections. The fraction $f$ is given by
Equation (\ref{ft}), $\alpha_0 = 0.008$, and $\xi_0 = 8$ for all the
tests. The dimensionless accretion rate $\dm$ is shown in the upper right
or lower right corner of each panel.}
\label{fig:xandj}
\end{figure*}

Figure (\ref{fig:xandj}) shows the resulting X-ray light curves for
four different accretion rates. It is worthwhile comparing Figure
(\ref{fig:xandj}) with Figure (\ref{fig:lcurves}), in which the plasma
ejections and random fluctuations in $f$ were absent.  The case with
the accretion rate just above $\dm_0$ (i.e., panel c in
Fig. \ref{fig:lcurves}) has not been affected as strongly as that with
the higher accretion rate; this is of course due to the fact that
little energy is ejected in the jet in the former case compared with
the latter. Yet panels (a) and (c) do look more chaotic, which is
entirely due to the fluctuating part in the fraction $f$ (see
Eq. (\ref{ft}).

\begin{figure*}[t]
\centerline{\psfig{file=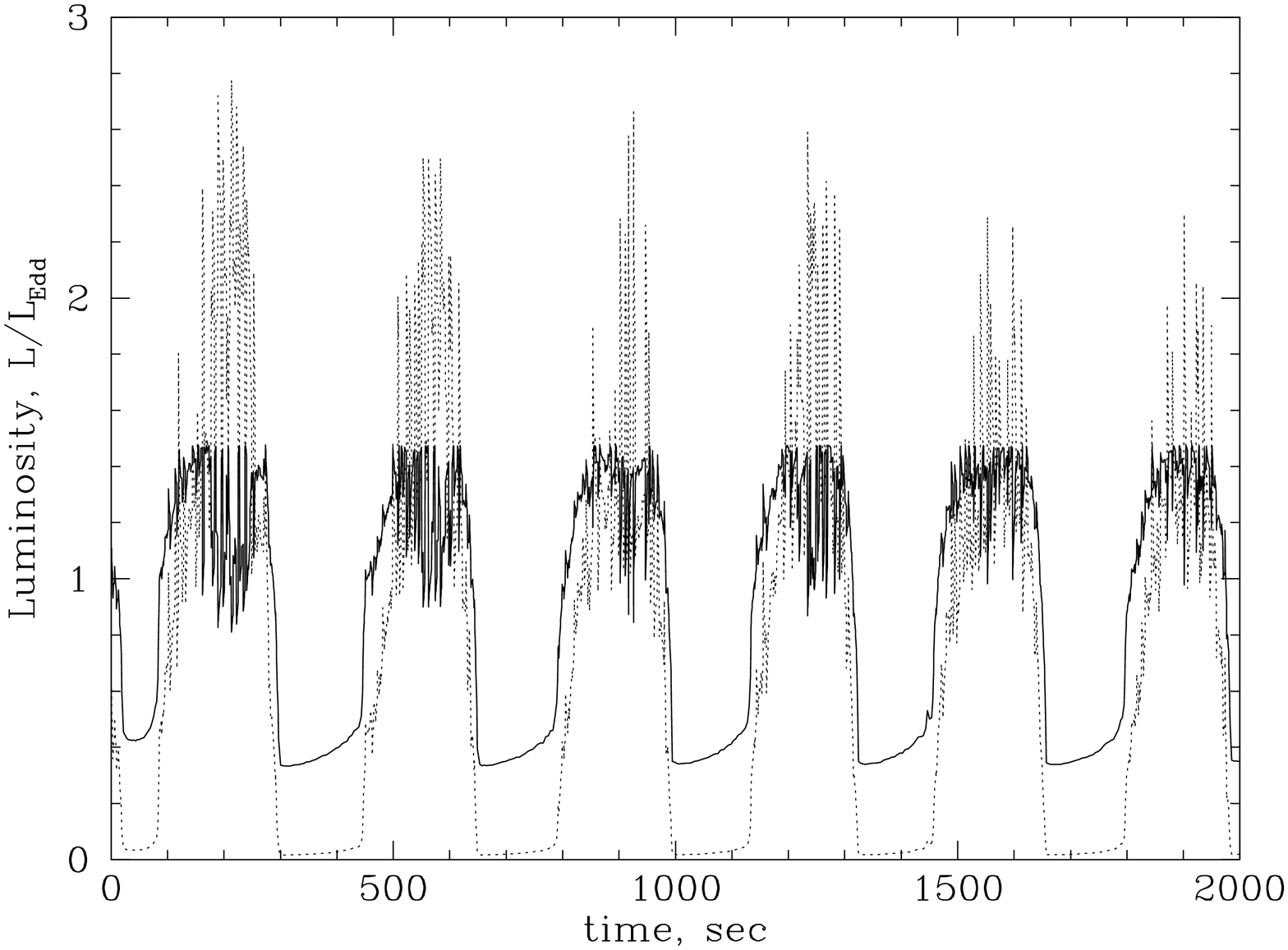,width=.65\textwidth,angle=-90}}
\caption{The X-ray luminosity (solid) and plasma ejection power (dotted)
corresponding to the $\dm=1.2$ panel of Figure (\ref{fig:xandj}).}
\label{fig:jet1}
\end{figure*}

Analyzing panels (b) and (d) of this figure, one notices that it turns
out to be possible to reproduce both the sharp square-wave like
pattern (c) and the ``inverted'' pattern (d) of Figure 1 with the {\em
same} prescription for $\eta_j$. It is the same process, i.e., the
plasma ejection, that allows our model to better reproduce the data
(panels [c] and [d] of Fig. 1); the only difference being that in
panel (d) the ejected fraction of the disk energy is much greater than
in panel (b). For the completeness of presentation, we also plot the
jet power $L_j$ with a dotted curve in Fig. (\ref{fig:jet1})
corresponding to the simulation presented in panel (b) of
Fig. (\ref{fig:xandj}).

Given the relatively good agreement between the model and the
observations of GRS~1915+105, we believe that we now have a better
handle on the key characteristics of accretion in this source. While
we feel less confident about the origin of the S-curve in the
$\Sigma-T_{\rm eff}$ space, we think that the geometry of the
accretion flow in GRS~1915+105 is that of a geometrically thin and
optically thick flow (see \S \ref{sect:geometry}). Most of the X-ray
luminosity is produced in an optically thin corona, that can cover the
entire inner disk or consist of localized transient magnetic flares
(e.g., Haardt, Maraschi \& Ghisellini 1994; Nayakshin \& Melia 1997;
Nayakshin 1998). In the latter case, the fraction $f$ of the power
transported from the disk into the corona and the jet must be thought
of as a time averaged variable.  Further, we believe that plasma
ejection events must be an integral part of the accretion process, and
are likely to be the result of an excessively large radiation
pressure. We will discuss this issue in a broader context in
\S \ref{sect:discussion}.

\section{Comparison to Previous Work}\label{sect:previous}

\begin{figure*}[t]
\centerline{\psfig{file=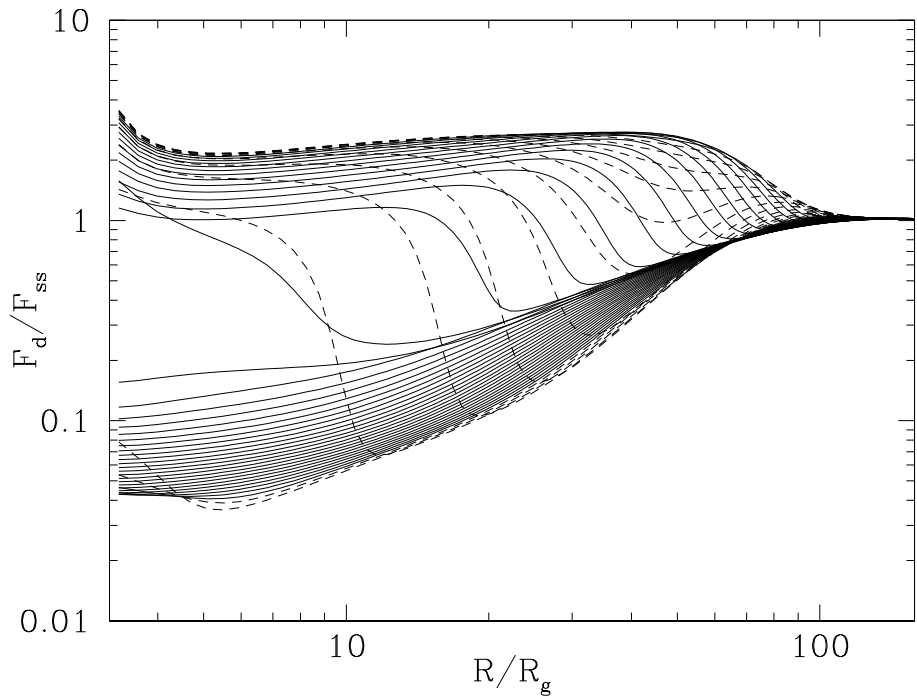,width=.65\textwidth,angle=-90}}
\caption{The ratio of the local X-ray luminosity to the luminosity
of a standard Shakura-Sunyaev disk with the given accretion rate
through one complete cycle. The model and its parameters are those
used to generate Figure \ref{fig:temp_tau_distr} (the disk behavior in
the tests presented in Fig. \ref{fig:xandj} is similar to the one
shown here, except for appropriate changes in the extent of the
unstable region in the disk). Note that the inner part of the disk can
indeed be very dim compared to its standard value and thus appear to
vanish below observability.}
\label{fig:dim}
\end{figure*}

Belloni et al. (1997a,b) were the first to fit the spectrum in
GRS~1915+105 with a two-component model, consisting of a
multi-temperature disk-blackbody with a variable inner disk radius
plus a power-law. Very importantly, these authors found that most of
the complex spectral variability can be explained by a rapid change in
the disk inner radius $R_{\rm in}$ and a corresponding change in the
temperature of the disk at $R_{\rm in}$. They have also shown a
correlation between the disk filling time scale and the inferred
radius of the inner disk.

Our results are qualitatively similar to the framework of the disk
instability discussed by Belloni et al. for GRS~1915+105 since in the
low state the inner disk is quite dim (up to a factor of 10-30, see
Figure \ref{fig:dim}) compared with what it should have been if the
disk were stable. Thus, the inner part of the disk can be largely
unobservable and can be said to be ``missing''.  At the same time, in
the high state, the inner disk is brighter than its equilibrium
luminosity by typically a factor of a few, which may make the outer
part of the disk (that is stable and so has just the ``normal'' flux)
seem comparatively dim. Under these conditions, it may be non-trivial
to distinguish observationally between a disk where the temperature is
a continuous function of R ($T\propto R^{-3/4}$) and a disk where the
temperature is lower than expected in the outer part of the disk,
since the outer disk carries a small fraction of the overall
luminosity.  In addition, the inner disk is never completely empty of
mass in our model, although the difference between the inner disk
surface density in the high versus that in the low state can be a
factor of $\sim$ few to 10 (see Fig. \ref{fig:temp_tau_distr}c).

Note that $R_{\rm in}$ as defined by Belloni et al. should be
identified with the largest radius reached by the heating wave in our
simulations. More specifically, during the high state, $R_{\rm in}$
should be defined as the radius of innermost stable orbit, whereas
during the low state it is approximately equal to the largest radius
in the disk in which $\pr = \pg$ (i.e., $R_{\rm in}\simeq 100$ for
Fig. \ref{fig:dim}).  Our values of $R_{\rm in}$ are in general much
larger than those obtained by Belloni et al.  The value of $R_{\rm
in}$ inferred from the observations could be larger if one were to fit
the GRS~1915+105 emission with a more complicated (and likely more
realistic) spectrum than a multi- temperature blackbody appropriate
for a Shakura-Sunyaev disk. Moreover, for the high accretion rates
observed in this source, the disk may become effectively optically
thin and thus radiate as a modified (rather than a pure) blackbody
(see discussion in Taam et al. 1997).  Since blackbody emission
produces the maximum flux for a given temperature, the modified
emission would require a larger $R_{\rm in}$. Finally, the values of
$R_{\rm in}$ obtained by us would presumably be smaller for a Kerr
black hole. We plan to address this question in the future.

Belloni et al. (1997b) showed that the outburst duration is
proportional to the duration of the preceding quiescent state, but
there is no correlation between the former and the quiescent time
after the burst. Profiles of outbursts in our model are rather
regular, so we believe we see a correlation among these three
quantities in contrast to the Belloni et al. results. On the other
hand, in many observations (other than the one presented in Belloni et
al. 1997b) GRS~1915+105 does not show a good correlation between the
duration of the outburst and the preceding quiescent phase
(T. Belloni 1999, private communication).

\section{Discussion}\label{sect:discussion}

We have systematically analyzed the physical principles underlying the
behavior of GRS~1915+105, and have arrived at several important
conclusions regarding the nature of the time-dependent accretion flow
in this system.  As discussed in Appendix
\S \ref{sect:geometry}, geometrically thick advection-dominated 
flows are unacceptable for this source, since they would produce
burst-like instabilities with a very small duty cycle compared to
those seen in the data, for accretion rates smaller than the Eddington
value.  To produce outbursts as long as $\sim 1000$ seconds,
geometrically thick accretion flows must have implausibly small values
of $\alpha$. However, if the accretion rate is highly super-Eddington,
then a thick ADAF disk could yield outbursts with a reasonable duty
cycle ($\sim 0.5$). But since the viscous time scale for geometrically
thick disks is as short as the thermal one, their light curves would
be rather smooth which again contradicts the data (see, e.g., panels c
\& d of Fig. 1).  It is also unclear how the observation (e.g.,
Belloni et al. 1997a,b; Muno, Morgan \& Remillard 1999) that the disk
extends down to the innermost stable orbit during the high state can
be reconciled with the structure of a geometrically thick, hot and
optically thin flow.

For these reasons, the standard cold accretion disk with a modified
viscosity law, a corona, and plasma ejections seems to be the only
reasonable choice. As we have shown in this paper, this geometry
indeed allows one to obtain rather good general agreement between the
theory and observations, and to understand particular features of the
light curves in terms of fundamental physical processes.

Nevertheless, a critical reader may question whether the relatively
large number of parameters introduced by us in this work allows us to
pin-point actual values of these parameters. We believe that the
answer to this question is ``yes'', because spectral constraints, used
very sparingly here, may prove to be quite restrictive in future
studies. For example, we have shown only the total luminosity light
curves in this paper, whereas the data also contain a wealth of
information about the spectral evolution in GRS~1915+105 (Belloni et
al. 1997; Muno et al. 1999). If future observations also provide more
examples of the disk-jet connection (e.g., Eikenberry et al. 1998),
then we have a means of constraining the dependence of both $\eta_j$
and the total fraction $f$ on the disk luminosity. Further, one should
also attempt to explore the fact that other transient and persistent
black hole sources do not exhibit instabilities similar to
GRS~1915+105, at least not to the same extreme degree. Thus, putting
all these observational and theoretical constraints together may be
quite effective in limiting the theoretical possibilities for the
instability and the coronal and jet activity.

As an example of the need for a detailed study of the spectra and QPOs
in the context of our model, we point out the following.  From the
results of Muno et al. (1999) it appears that QPOs are present when
the power-law component dominates the spectrum, which usually happens
in the low state (``low'' means the lower count rate). Further, these
authors find that for the burst profiles similar to the one shown in
Fig. 1(d), the QPOs are present during the ``high'' state, which is
opposite to most of the other cases. If QPOs indeed track the presence
of a vigorous corona, this would imply that although our panel (d) in
Fig. (\ref{fig:xandj}) looks similar to Fig. 1(d), it does not
represent the actual situation very well. Namely, the data seem to
imply that the higher count rate states actually have
lower accretion rates through the inner disk than the lower luminosity states
(because excess luminosity escapes into the jet). It is thus
possible that plasma ejections occur during the sharp dips in Fig 1(d)
rather than during the longer phases (e.g., from $\sim 420$ to $\sim$
960 sec in this figure).  The ``M-shape'' of the outbursts in
Fig. (\ref{fig:data}d) may then be understood because these would be
the low states as tracked by the accretion rate, and the low states of
the three other panels (a-c) of Fig. (\ref{fig:data}) show similar
shapes albeit with lower count rates. Although we have not yet made a
detailed study of this suggestion, we can probably accommodate it
within our model, but only if we allow the parameters in our viscosity
law to depend on the disk luminosity. {\em Note, however, that in
either case we must require plasma ejections from the inner disk in
order to understand panel (d) of Fig. (1).}

The discussion above of course bears on the particular choice of the
viscosity law that we have made in this paper. We emphasize that our
model is still rather empirical and we have not identified the actual
physics of the viscosity law in GRS~1915+105. The real physics might
well be yet more complex, and in fact, the existence of the S-curve
might be due to other than radiation pressure instabilities, perhaps
involving an effect that we do not currently understand. However,
based on the results presented in this paper, we feel confident that
there is an S-curve in this source, and that this curve can be modeled
approximately by the viscosity law given by equation (\ref{eq3}).

A clear shortcoming of our work is the use of a non-relativistic disk
around a non-spinning black hole. We plan to improve this in our
future work. We expect that this will change the ``reasonable'' values
of the parameters that we used, such as the fraction $f$, and the
inner disk radius, of course, but that it will not affect the general
nature of our conclusions.

\acknowledgments{}

The authors thank T. Belloni, R. Taam, M. Muno, E. Morgan,
R. Remillard, E. Vishniac and D. Kazanas for useful discussions.  This
work was supported in part by NASA Grants NAG5-8239 and NAG5-4057,
under the Astrophysics Theory Program.

\appendix{

\section{A. On the geometry of the accretion flow in GRS~1915+105}
\label{sect:geometry}

In almost any accretion disk theory, the ratio of the disk half
thickness $H$ to the radius $R$ becomes comparable to unity when the
accretion rate approaches the Eddington limit, which is almost
certainly the case for this source. However, the data strongly suggest
that $H/R\ll 1$ in GRS~1915+105. Indeed, a state transition in the
unstable portion of the disk cannot occur faster than one thermal time
scale, since the gas needs to be heated up or cooled down
substantially to go from one branch of the solution to the other. Let
us now consider panel (c) or (d) of Figure 1, since these are the
highest luminosity cases, where $H/R$ should be the largest and so
time variability constraints are the strongest there. By looking at
these two panels, one notices that the observations require that the
thermal time scale be smaller than about 10 seconds ($t_{\rm th}\simlt
10$ sec), whereas the cycle time is of the order of a thousand
seconds.  The cycle time should be associated with the disk viscous
time scale, i.e., the time during which the disk surface density
$\Sigma$ can change appreciably (and so cause a state transition),
which is $t_{\rm visc}\sim (R/H)^2 t_{\rm th}$ (see \S 5.8 of Frank et
al. 1992). This implies $(H/R)\simlt 0.1$

\subsection{ADAFs are ruled out for GRS~1915+105}\label{sect:adaf}

ADAF configurations (see, e.g., Narayan \& Yi 1995) have $H/R\simeq 1$
even for very low accretion rates. For this reason, ADAFs are ruled
out for GRS~1915+105. Since by definition the thermal time scale in
ADAFs is longer than the viscous time scale, the rise and fall in the
light curve produced by an ADAF always would be very gradual and of
the same duration as the whole outburst phase. More specifically, an
ADAF is like a slim disk, in the sense that it can produce short spiky
bursts, because $t_{\rm visc}$ in the high state can be quite short
(see Figures 7-9 in Szuszkiewicz \& Miller 1998), but it cannot
explain {\em long outburst cycles with very fast transitions}, such as
the ones seen in panels (c,d) of Figure 1. Furthermore, the viscous
time is exceedingly short for disks where advection is important:
$t_{\rm visc} \sim \alpha^{-1} t_{\rm dyn}\sim 3\times 10^{-3} M_1
\alpha^{-1}$ sec, where $M_1$ is the black hole mass in units of
$\msun$. This numerical value of the viscous time scale corresponds to
$r\equiv R/R_g = 10$ (since this appears to be the appropriate value
based on observations by, e.g., Belloni et al. 1997a,b and Muno et
al. 1999).  Thus, to reconcile this with the observed viscous time
scale of $\sim 10^3$ sec, one should expect a very small
$\alpha$-parameter, i.e., $\alpha\sim 3\times 10^{-6}$, which seems
rather unrealistic. Moreover, even if one {\em can} neglect all the
interactions between electrons and protons except for Coulomb
collisions (otherwise ADAFs would never form), as well as omit the
Ohmic heating of electrons (see Bisnovatiy-Kogan \& Lovelace 1998;
Quataert 1998), one still has the constraint that ADAFs can exist only
below $\dm \simlt \alpha^2$ (see Narayan \& Yi 1995), which is clearly
too small for GRS1915 if $\alpha\sim 3\times 10^{-6}$.

\subsection{Hot central region}\label{sect:hot_ball}

An alternative model that has been suggested for generating X-rays in
a black hole accretion disk invokes the presence of a central
quasi-spherical emission region that is optically thin and hot. The
main appeal of this type of model is its geometrical simplicity, which
however ends immediately when one attempts to supplement this picture
with realistic physics. We are unaware of any detailed and successful
attempt to model this central source (i.e., with the inclusion of
energy and pressure balance for the hot plasma), except for ADAFs and
the two-temperature model for Cyg X-1 (Shapiro, Lightman \& Eardley
1976). The first model is unacceptable for the reasons discussed in \S
\ref{sect:adaf}, whereas the second model is known to be thermally
unstable (and does not contain the two upper and lower stable
solutions needed for GRS~1915+105) and thus its fate is uncertain. In
addition, Misra \& Melia (1996) showed that in the context of the
two-temperature model, there should also be a rather extended
transition region between the hot source and the cold outer disk. The
requirement of the existence of such a transition region in
GRS~1915+105 appears to contradict the results of Belloni et
al. (1997a,b), who were able to fit the source spectra reasonably well
with a simple multi-color disk blackbody model using a rather small
$R_{\rm in}$.

Taam et al. (1997) and Chen et al. (1997) argued that the geometry of
the disk in GRS~1915+105 is that of a cold outer disk extending from
some $r_{\rm in} \sim 30$ to infinity, plus a hot, optically thin
inner region between $r = 3$ and $r_{\rm in}$. They interpreted the
spectral changes from hard/lower luminosity states to softer/higher
luminosity states as arising from a decrease in the value of $r_{\rm
in}$. The inner disk region was not modeled by these authors, and
instead it was assumed that its behavior is simply modulated by the
changes in the accretion rate through the outer disk.  But let us now consider this suggestion in 
more detail. Suppose that when the accretion rate increases, the
inner region shrinks in size from $r_c$ to $r_h< r_c$. 
Since the spectral index changes considerably
from the cold to the hot state, $r_c-r_h\sim r_c$, i.e., the inner
disk radius should decrease substantially (as an illustration of our
point, see Poutanen, Krolik, \& Ryde 1997 for the case of spectral
transitions in Cyg X-1, where it is found that $r_c\simeq 40$ and
$r_h\simeq 8$). The region $r_h < r < r_c$ should cool to the
temperature and height appropriate for the Shakura-Sunyaev disk. To be
applicable to GRS~1915+105, this cooling must happen on a thermal 
time scale. Thus, $\Sigma\sim
\Sigma_0$, where $\Sigma_0\simlt $ few is the hot inner disk column
density. Let us now estimate the accretion rate through this region
immediately after a transition takes place. The accretion rate is
given by $2 \pi R\Sigma v_R$, and it is only $v_R$ in this expression
that can change on a thermal time scale. From Equation (\ref{vr}) we
have $v_R\sim 3 \alpha c r^{-1/2} (H/R)^2$.  The ratio
of the accretion rate in the high state, $\dot{M}_h$, to that in the
cold state, $\dot{M}_c$, is
\begin{equation}
{\dot{M}_h\over \dot{M}_c}= \left({H_{\rm cold}\over H_{\rm
hot}}\right)^2
\sim \left( {H_{\rm cold}\over R}\right)^2 \sim 10^{-3}\;,
\label{nonsense}
\end{equation}
where a value of $H_{\rm cold}/R\sim 0.03$ is used, since this is when
the radiation to gas-pressure transitions happen in the standard
accretion disk theory for $M_1\sim 1$ (e.g., SZ94). The inner hot
region is starved and its luminosity is correspondingly lower (by a
factor of $10^3$ if there is no advection, or by a factor of $10^6$ if
advection is important, since in the latter case the efficiency of
conversion of the accretion power into radiation decreases
proportionally to the decrease in $\dot{M}$). Furthermore, the
luminosity of the torus $r_h < r < r_c$ also decreases in the same
proportion, because the local flux from the disk is $F_d = 3GM\dot{M}/
8\pi R^3 \propto \dot{M}$. We conclude that if the radius of the hot
inner region suddenly increases, then the {\em luminosity of the
system should first significantly decrease before it goes up}.

The time necessary for the inner disk to recover from this and become
brighter than it was in the cold (hard) state is simply given by the
filling time of the transition region. This time is $t_r\sim (R_c -
R_h)/\nu \sim R_c/\nu\equiv t_v(R_c)$, where $t_v(R_c)$ is the viscous
time at $R_c$. The disk instability itself in the Taam et al. (1997)
picture happens at a yet larger radius. However, the results of
Belloni et al. (1997a,b) show that this radius cannot be much larger
than $r_c$. In that case, we have just shown that the cold to hot
transition in the framework of a central varying source should first
have a dip in luminosity and a slower recovering phase, with the rise
time comparable to that of the whole instability cycle.  It is also
worthwhile mentioning that following this reasoning, a transition from
the hot to the cold state would first produce a strong flare. The
problem here is that one needs to dispose of the transition region
mass during a thermal time scale in order to extend the hot region
from $r_h$ to $r_c$, which means that during that time the accretion
rate exceeds the average accretion rate by approximately
$(R/H_c)^2\sim 10^3$. Needless to say, neither substantial dips nor
bright flare-ups are observed in GRS~1915+105 at the end of the hot
state, and for this reason we believe that the geometry of a central
source with a varying radius is inconsistent with the observations.

In addition, it is questionable whether the boundary conditions used
by Abramowicz et al. (1995) at the interface between the cold and hot
portions of the disk (i.e., $dT/dR = 0$ and $d\Sigma/dR = 0$) are
justifiable, because at this point one would expect these gradients to
be the largest in the disk. It appears to us that more appropriate
boundary conditions would be given by the continuity of mass flow at
$r=r_{\rm c}$, i.e., $d\dot{M}/dr = 0$, and continuity in the radial
energy flux, or specifying a temperature gradient at this point,
$dT/dr = -(T_{\rm in}- T_{\rm c})/\delta l$, where $T_{\rm in}$ is the
inner disk temperature and $\delta l$ is a characteristic scale of the
``boundary layer'' between the two disk segments. Of course, a
completely self-consistent solution would only be obtained when one
also includes the time-dependent inner disk equations and solves the
whole disk evolution  through the instability.

Concluding this section, we believe that the geometry of a hot central
source plus a cold disk with a variable interface between the two is
not appropriate for GRS~1915+105, and that one needs to do a careful
study to see whether this model can be applied to any time-dependent
phenomena in GBHCs or LMXBs. Further, a central source with a fixed
boundary is not ruled out by our considerations here, but then the
main attraction of the model -- the simple picture of spectral
hardening or softening as being due to change in $r_{\rm in}$ -- no
longer exists.

\subsection{Cold Disk Plus a Hot Overlying Corona}
\label{sect:corona}

A geometry that satisfies the constraints $H/R\simlt 0.1$ is that of a
geometrically thin and cold (compared with an ADAF, for example)
accretion disk with a hot, optically thin corona above it. The
accretion disk height scale $H$ is reduced from the value given by the
Shakura-Sunyaev theory due to the addional coronal cooling by the
factor $1-f$ (see Equation 7 in Svensson \& Zdziarski 1994).  In this
picture, the thermal disk emission corresponds to the
multi-temperature blackbody emission modeled by Belloni et al.
(1997a,b) and Muno et al. (1999), whereas the non-thermal power-law is
generated within the corona. The corona itself may be localized,
consisting of magnetic flares (e.g., Haardt, Maraschi \& Ghisellini
1994; Nayakshin 1998) or it may be extended, with a vertical scale
anywhere between $H$ and the local disk radius. This latter
possibility does not contradict our previous estimates of $t_{\rm
th}/t_{\rm visc}$, since it is the cold disk below the corona wherein
most of the mass is contained, and thus it is the scale height of the
cold disk that defines $t_{\rm visc}/t_{\rm visc}$.

\section{B. Effects of the Enhanced Energy Transport}\label{sect:f}

Throughout our studies, we deliberately used the fraction $f$ in the
cooling term in the energy balance equation (eq. \ref{eq5}), rather
than in the heating term, as was done by Abramowicz et al. (1995) and
Taam, Chen \& Swank (1997).  The latter approach assumes that the
energy released in the disk is reduced by the factor $(1-f)$, i.e., a
fraction $(1-f)$ of the viscously released energy was not even
deposited in the disk, and that the cooling rate is the same as in the
standard model. In our approach, all the energy is released in the
disk, and it is the energy transport out of the disk that is being
increased rather than the heating being decreased. We believe that our
formulation of the problem follows what physically occurs in a
disk-corona system. In static accretion disk models, wherein one can
neglect radial energy transport, both formulations will give identical
results, since only the two first terms on the right hand side of
Equation (\ref{eq5}) are non zero. However, in a time dependent
situation, moving the factor $(1-f)$ from the cooling term to the
heating term results in an unphysical re-normalization of these two
terms relative to the other terms of Equation (\ref{eq5}), and thus
may give rise to a serious error.

\begin{figure*}[t]
\centerline{\psfig{file=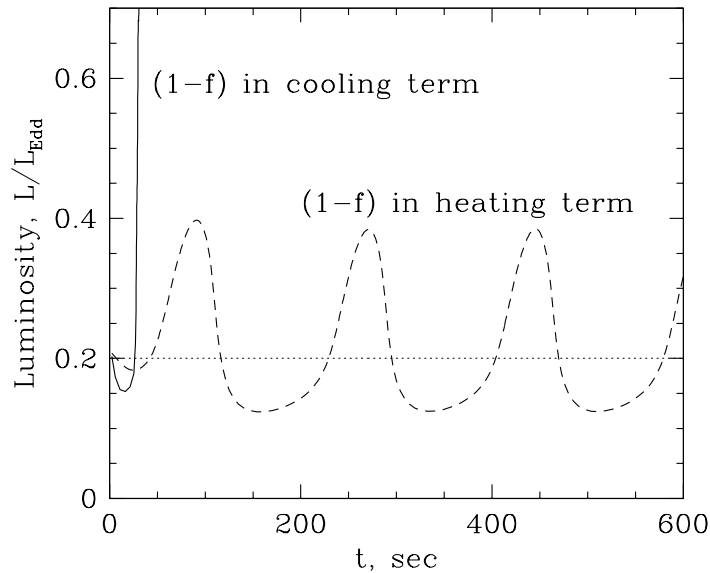,width=.5\textwidth,angle=0}}
\caption{Instabilities of the standard accretion disk theory with the
coronal cooling included in the (i): cooling term, as is done in this
paper (solid line); (ii) heating term, which is unphysical (dashed
line). Note that the first model would actually saturate at a very
high luminosity, but the resulting pattern of outbursts would consist
of a series of spikes or sharp bursts, and would never be able to
reproduce variability such as that seen in panel (c) of Figure 1
for GRS~1915+105.  The model parameters used to generate this figure are:
$\dm = 0.2$, $f = 0.75$, and $M = 10\msun$.}
\label{fig:compare2}
\end{figure*}

We have run our disk evolution code for different values of the
accretion rate $\dm$ and the fraction $f$ to determine whether a
supplemental vertical energy transport added to the standard
Shakura-Sunyaev disks with constant $\alpha$ can explain the
observations of GRS1915. All of our models with accretion rates
sufficiently high to produce radiation pressure dominated regions over
a portion of the standard disk were found to be unstable.  Moreover,
none of the simulated bolometric light curves became saturated, i.e.,
all the models ran away to arbitrarily high temperatures and low
optical depths. Physically, this means, of course, that we would need
to modify our code to include the optically thin case; this would
presumably lead to a saturation in the instability, followed by a
gradual decline in the luminosity and a return to the gas-dominated
quasi-stable state.  However, such oscillations would be fundamentally
different than the ones observed in GRS~1915+105. In particular, the
standard disk (with $\alpha =$ const) plus corona system would produce
quasi-periodic spike-like outbursts, with a very large ratio of
maximum to minimum luminosity, and a very small duty cycle (see, e.g.,
Cannizzo 1996, 1997), whereas GRS~1915+105 displays a variety of
behaviors, many of which have duty cycles larger than a half.  Thus,
we conclude that the standard viscosity law, even with the addition of
a corona cannot explain the observations of GRS~1915+105.

On the other hand, Abramowicz et al. (1995), and Taam et al. (1997) 
found a mild oscillatory behavior for a range of $\dm$ and
$f$. We believe that the disagreement is entirely due to the
difference discussed above in the treatment of the effects of the
corona above the disk. To check this, we placed the {\it cooling} term
due to the additional energy transfer from the disk to the corona,
into the heating term, as was done by these authors, and ran several
tests. With this, we were indeed able to reproduce their results. In
Figure (\ref{fig:compare2}), we show the difference in the resulting
light curves that the positioning of the coronal cooling term makes.
The solid curve shows the results of our code with $f$ placed in the
cooling term, as explained above. In contrast, the dashed curve shows
the light curve with the coronal cooling placed in the heating term.
It can be seen that the instability indeed saturates when one uses the
latter (unphysical) formulation of the problem, and oscillations with
a reasonable amplitude follow.

In all of these calculations, we have assumed that the fraction of
power transmitted by the additional energy transfer mechanism is a
constant independent of the local conditions in the disk. However, the
spectrum of GRS~1915+105 in the high state is generally (but not
always) softer than it is in the low state (e.g., Belloni et
al. 1997a,b; Muno et al. 1999), so that $f$ probably decreases from
the low to the high state. One then might wonder whether allowing $f$
to vary in accordance with the observations would change our
conclusions and allow the standard accretion disk theory to reproduce
the observed instabilities in GRS~1915+105. Unfortunately, this effect
actually goes the wrong way, making the instability stronger. The
presence of the corona always decreases the extent of the unstable
region in the disk (e.g., Svensson \& Zdziarski 1994). In order to
suppress the runaway process illustrated by the solid curve in
Fig. (\ref{fig:compare2}), one would need the fraction $f$ to be
larger in the high state than it is in the low state, which is
opposite to what is observed. Summarizing, we discount the notion that
a hot corona is sufficient to make the disk unstable in the correct
manner to explain the observations. The standard accretion disk with a
standard viscosity prescription, with or without a corona, is unstable
in the radiation-dominated regime and yields a burst-like behavior
that is incompatible with the observations of GRS~1915+105.

}

{}

\end{document}